%% file: template.tex
\documentclass{article}

\usepackage{arxiv}

\usepackage[utf8]{inputenc} 
\usepackage[T1]{fontenc}    
\usepackage{hyperref}       
\usepackage{url}            
\usepackage{booktabs}       
\usepackage{amsfonts}       
\usepackage{nicefrac}       
\usepackage{microtype}      
\usepackage{lipsum}
\usepackage{graphicx}

\usepackage[utf8]{inputenc}
\usepackage{enumerate}
\usepackage{graphicx}
\usepackage{inputenc}
\usepackage[linesnumbered,ruled,vlined]{algorithm2e}
\usepackage{amsmath}
\usepackage{algorithmic}
\usepackage{upquote}
\usepackage{xcolor}
\usepackage{comment}
\usepackage{subfigure}
\usepackage{mathtools}
\usepackage{multirow}
\usepackage{lscape}
\usepackage{fancyhdr}
\usepackage{tikz}
\usepackage{graphicx}
\usepackage{listings}
\usepackage{xcolor}
 \usepackage{mathtools}
 \usepackage{booktabs}

\definecolor{codegreen}{rgb}{0,0.6,0}
\definecolor{codegray}{rgb}{0.5,0.5,0.5}
\definecolor{codepurple}{rgb}{0.58,0,0.82}
\definecolor{backcolour}{rgb}{0.95,0.95,0.92}
 
\definecolor{lightgray}{gray}{0.8}
\newcommand\encircleblack[1]{%
	\tikz[baseline=(X.base)] 
	\node (X) [draw, shape=circle, scale=0.7, inner sep=0, fill=black, text=white] {\strut #1};%
}

\definecolor{lightgray}{gray}{1}
\newcommand\encirclewhite[1]{%
	\tikz[baseline=(X.base)] 
	\node (X) [draw, shape=circle, scale=0.7, inner sep=0, fill=white, text=black] {\strut #1};%
}

\graphicspath{ {./images/} }

\title{VAR-DRAM: Variation-Aware Framework for Efficient Dynamic Random Access Memory Design}

\author{ 
    {Kaustav Goswami} \\ 
	Department of Computer Science\\
	University of California, Davis\\
	California, United States \\
	\texttt{\href{mailto:kggoswami@ucdavis.edu}{kggoswami@ucdavis.edu}} \\
	\And
	{Hemanta Kumar Mondal} \\
	Department of Electronics and Communication Engineering\\
	National Institute of Technology Durgapur\\
	West Bengal, India \\
	\texttt{\href{mailto:hemanta.mondal@ece.nitdgp.ac.in}{hemanta.mondal@ece.nitdgp.ac.in}} \\
	\And
	{Shirshendu Das} \\
	Department of Computer Science and Engineering\\
	Indian Institute of Technology Ropar\\
	Punjab, India \\
	\texttt{\href{mailto:shirshendu@iitrpr.ac.in}{shirshendu@iitrpr.ac.in}} \\
	\And
	{Dip Sankar Banerjee} \\
	Department of Computer Science and Engineering\\
	Indian Institute of Technology Jodhpur\\
	Rajasthan, India \\
	\texttt{\href{mailto:dipsankarb@iitj.ac.in}{dipsankarb@iitj.ac.in}} \\
}
\begin{document}
\maketitle
\begin{abstract}
    Dynamic Random Access Memory (DRAM) is the \textit{de-facto} choice for main memory devices due to its cost-effectiveness. It offers a larger capacity and higher bandwidth compared to SRAM but is slower than the latter. With each passing generation, DRAMs are becoming denser. One of its side-effects is the deviation of nominal parameters: process, voltage, and temperature. DRAMs are often considered as the bottleneck of the system as it trades off performance with capacity. With such inherent limitations, further deviation from nominal specifications is undesired. In this paper, we investigate the impact of variations in conventional DRAM devices on the aspects of performance, reliability, and energy requirements. Based on this study, we model a variation-aware framework, called VAR-DRAM, targeted for modern-day DRAM devices. It provides enhanced power management by taking variations into account. VAR-DRAM ensures faster execution of programs as it internally remaps data from variation affected cells to normal cells and also ensures data preservation. On extensive experimentation, we find that VAR-DRAM achieves peak energy savings of up to 48.8\% with an average of 29.54\% on DDR4 memories while improving the access latency of the DRAM compared to a variation affected device by 7.4\%.
\end{abstract}


\input{Text/introv2}
\input{Text/motivation}
\input{Text/relatedv2}
\input{Text/background}

\input{Text/methodv2}
\input{Text/evaluation.tex}
\input{Text/analysis}
\input{Text/conclusion}


\bibliographystyle{unsrt}  
\bibliography{template.bib}  

\end{document}

%% file: Text/introv2.tex

\section{Introduction}
\label{sec:intro}

    In current computing systems, Dynamic Random Access Memory (DRAM) based main memories constitute a significant portion of the total power consumption of the system. For example, studies done such in~\cite{ware, hoelze, barosso} show that DRAM devices consume up to 40\% power in server-class systems and up to 50\% power in graphics cards. Towards lowering the power that is consumed by DRAMs, vendors have implemented different power-aware DRAMs designs such as DDRx memory~\cite{micro,ddr4}, LPDDRx (low power DDR)~\cite{microLPDDRx} etc.). In these aforementioned technologies, the reduction in power consumption is achieved ue to the advancement of CMOS process technologies. Subsequent reductions are possible through scaling the supply voltage to the core DRAM array as well as the peripheral circuitry. In other directions, the power consumed by DRAM can be reduced via scaling of the supply voltage, and putting the different DRAM banks to different power states. Such variations have been modeled in works~\cite{sarangi,process} to identify the decaying portions of the chip.

    The classic monolithic design of a typical DRAM has been enhanced to accommodate better performance in minimal power requirements. A modern-day DRAM device is organized into ranks and banks. Although this memory layer may not be active at all times, the capacitor-based chip has to be refreshed in order to maintain the integrity of the data it contains. It has been well established that ranks can be transitioned into \textit{low power mode}~\cite{micro, delaluz}. However, DRAM banks on the other hand, have been rarely exploited for such power-saving techniques. Few works have been done towards this~\cite{delaluz,delaluz-bank}, which remains to be exploited on a large scale. One of the prominent reasons for this is due to the fact that DRAM power is largely calculated on a per-rank basis~\cite{delaluz, micro, ddr4}. DRAM power consumption mechanism, later discussed in Section~\ref{subsec:dram-pow}, has always been centered mostly around a rank-wise approach. This opens up new opportunities for detailed investigations into the power efficiency of DRAM banks. The biggest beneficiary of such a power management technique would be LPDDR memories, installed on battery-operated devices like smartphones etc. However, modifications to the conventional architecture are necessary in order to exploit power savings from individual DRAM banks.

    Another problem faced towards recent advancements in technology is the scaling of chips. This has caused a deviation in the process parameters of the digital chip from its nominal specified values. This effect has greatly reduced the ability of uniform performance by a device. Sections of the device under-performs, thus reducing the total throughput of the system. The trend of enhancement is towards reduced chip sizes. With chip size dipping below 32 nm technology node, it has been reported that process, voltage, and temperature (PVT) show a variation from its nominal specifications~\cite{sarangi,process,ras_timing}. On the application's end, the requirements of both compute-intensive, as well as data-intensive programs, are increasing. With such vivid requirements, one cannot ignore effects like parameter variations. Such effects affect the desired performance. Affected devices are slow, and consume a higher amount of power than their estimated counterparts. This also jeopardizes the reliability of the data present in these chips as the reliability of the chips depends on the process conditions. Variation is induced by several fundamental effects and is a combination of systematic and random effects. While systematic effects include lithographic lens aberrations, random effects include dopant density fluctuation~\cite{sarangi}.

    These aforementioned variations give significant opportunities for optimization which motivates our work. Previous works have shown variation prone DRAM devices affect access latency~\cite{process,ras_timing}, retention time~\cite{raidr}, and reliability of the data. We propose a technique to exploit power savings from these variations affected DRAM cells. We observe that powering down these affected cells results in a significant amount of energy savings. These cells, therefore, become ideal candidates for powering down in order to obtain energy savings. In a DRAM device, the last level parallelism is offered at the DRAM bank level. Therefore, we select to switch power from DRAM banks. Another challenge while working with bank switching at regular intervals is that the data contained in these banks which are prone to get lost. Under such circumstances, preserving the data, or the \textit{state} of the DRAM, leads to having both advantages as well as disadvantages. The advantages are that the data need not be re-fetched or re-calculated once the banks are re-opened for use, and subsequent operations can continue in a seamless manner. The major disadvantage of preserving the state is the fact that there exists performance overheads, which may potentially decay the performance on the end application. Several works have proposed techniques where the state is both preserved and destroyed. In our work, we propose a technique that provides state preservation via a very low overhead data structure. The implementation of the data structure incurs low overhead on both performance and additional area consumed.

%% file: Text/motivation.tex
\subsection{Motivation and Contribution}

    Our primary aim is to mitigate variation related challenges that we face in normal DRAM memory. Once variation affected areas are identified, we use an enhanced power management mechanism to power down these under-performing banks while remapping the data present in these cells to normally functioning banks thus maintaining access latency and ensuring higher reliability. The conventional memory controller is modified to account for variation data. The controller now transitions these banks into a new \textit{ultra-low-power mode} using a power gated circuit for DRAM banks, which we refer to as simply \textit{powering down} of banks. As for preserving the data of these powered down banks, we remap both addresses and their respective data present. The remapping logic is flexible, thus allowing us to use it for other applications, which is later discussed in the paper. VAR-DRAM saves up to 48.8\% of DRAM energy, averaging at 29.54\%. VAR-DRAM is a \textit{framework}, where we are potentially able to stack other energy-saving mechanisms to achieve higher energy savings while maintaining a nominal access latency. To verify our proposed technique, we have validated our proposed technique extensively. Collectively, our work provides the following concrete contributions:
    
    \subsubsection{An efficient power management mechanism}
        Prior works are done towards bank-level power management does include bank-wise refresh~\cite{microLPDDRx} and other software-based power management techniques~\cite{delaluz-bank}. However, powering down DRAM banks using power gating is relatively new for a DRAM design. In addition, it is also sparsely explored. Also, we investigate methods for saving power at the bank-level from variation affected under-performing components. This becomes particularly essential in context of battery-operated mobile devices. Lowering power consumption would enable longer battery drain times.
        
    \subsubsection{Memory devices with low latency} 
        Variations affect the access latency of the memory devices. VAR-DRAM presents a variation-aware address remapping, which favors normally working DRAM cells over variation affected ones, thus reducing the access latency of the memory device.
        
    \subsubsection{A lightweight address remapping logic} 
        We use a search and space-efficient data structure for remapping called \textit{trie}, which provides better space complexity than tables with logarithmic lookup times. Maintaining it eliminates re-computation costs, as the state of the DRAM is preserved throughout, but possesses an additional storage overhead. Our findings (Section~\ref{sec:result}) clearly show that this overhead is marginal. Moreover, the remapping logic can also be used for remapping weak rows of a DRAM device, minimizing the blocking time of the DRAM device during DRAM refreshes.

    The paper is organized as follows. Section~\ref{sec:related} provides details of prior energy-efficient DRAM-related works. Section~\ref{sec:background} then gives an introduction to DRAM devices, variations, and, how DRAM cells are affected by it. Section \ref{sec:method} gives an intrinsic detail on the implementation technique. Section~\ref{sec:eval} lays the foundation of the simulator platform used to evaluate and subsequently discusses the experiments conducted. Section~\ref{sec:result} sheds a light on the results obtained and provides an analysis for the same. Finally, Section \ref{sec:conclusion} concludes this work with scope for improvements and extensions in the future.

%% file: Text/relatedv2.tex
\section{Related Works}
	\label{sec:related}
    Exploring power savings in DRAM devices is one of the key techniques employed in power-efficient systems. Studies such as~\cite{ware, hoelze} report that DRAM devices consume a significant portion of the system's power. A significant number of prior works have proposed towards saving power for DRAM devices. Delaluz et al.~\cite{delaluz} exploited the prolonged behavior of idle ranks for saving power. Lebeck et al.~\cite{lebeck} demonstrated how a DRAM rank can be transitioned into low power mode. They proposed a method to cluster DRAM accesses into some specific chips so that idle chips can be transitioned into low power mode. This concept of powering down ranks is then established as a standard technique to save power from DRAM devices. Other techniques based on schedulers were then studied in works including~\cite{lebeck,delaluz,delaluz-kande}. The authors in~\cite{delaluz-kande} proposed a method to dynamically map addresses into some specific set of banks in order to exploit power savings. 
    Hassan et al.~\cite{crow} proposed an In-DRAM cache for DRAM-based memory systems. The proposed model makes a copy of a DRAM row on the cache. Later, during the execution of the program, if the same row is referenced, it is activated simultaneously from both the primary array and, as well as from the In-DRAM cache (also known as CROW-cache). Another aspect of saving power in DRAM devices is centered around periodic DRAM refreshes. In the work~\cite{raidr}, the authors propose a selective refresh on variation affected cells in order to maintain the data properly and eliminate unnecessary refreshes. Bandwidth-aware power management~\cite{10.1109,liu2010} for higher DRAM page hit rate is another technique to reduce the power requirements of DRAM devices. Lee et al.~\cite{leecho} proposed a bandwidth-aware page migration technique for heterogeneous memories.
	
    Shutting down DRAM banks or transitioning DRAM banks into low power mode is a topic with limited study. Prior works on this aspect include~\cite{delaluz-bank,isqed}, wherein the former, authors propose both software and hardware-based techniques to exploit power savings from DRAM banks. The latter attempts to close under-utilized DRAM banks in order to save power. Another work~\cite{bank-sensitive-model} includes a bank-sensitive power model for a DRAM power simulator called DRAMPower~\cite{drampower}. The existing implementation of a DRAM module is biased towards power management in a rank-wise manner~\cite{micro,ddr4}. Low power memory technologies like LPDDR DRAM have a provision of a bank-wise refresh instead of an all-bank refresh~\cite{microLPDDRx}. The DRAM simulator called DRAMSim2~\cite{dramsim} also implements a heuristics-based technique to demonstrate low power mode in DRAM systems during simulations. Micron Technologies implements a low power mode for idle ranks for their DDR3 DRAM system~\cite{micro} where no I/O accesses are allowed to the rank which results in lower consumption of power. 

    The fabrication process of manufacturing chips is susceptible to variations~\cite{process,sarangi}. These variations lead to the aberration of nominal parameters of Process (P), Voltage (V), and Temperature (T) of these chips. Variation is a well-studied topic for DRAM devices. Hamamoto et al. investigated DRAM data retention distribution in their work~\cite{o34}. They reported that boron concentration in the p-well of memory cells characterizes the data retention distribution. Further, variation affects the reliability of the data present on the device which plays an important role in real-time systems and secured systems where corrupt data can lead to catastrophic consequences. Emerging memories and non-volatile memory technologies are gaining momentum in development as these provide a feasible alternative to DRAM-based memories~\cite{nvm}. However, these technologies also have variations~\cite{stt-pv}. Even flash storage devices like SSD are affected by PV~\cite{other_paper}. 

	Access latency of a DRAM device is affected by PV~\cite{sarangi, process,ras_timing}. Zhao et al.~\cite{process} studied the effects of PV on a DRAM-based last level cache (LLC) where they tried to mitigate slow memory accesses by migrating the data to some other locations of the LLC so that the access latency is maintained uniformly. Sarangi et al.~\cite{sarangi} proposed a statistical method called \textit{VARIUS}, where the decaying portions of a chip can be identified. A study by Ghose et al.~\cite{vampire} concludes that most off-the-shelf DRAM devices are prone to variation. The authors did a study on each granularity of the DRAM device and compared it with state-of-the-art simulator results and concluded that there exists a large difference between real and simulated hardware results. Current DDRx DRAM systems perform row migrations in order to mitigate the issue of PV-affected defective rows~\cite{sample}. Although this mechanism does mitigate the issue of reliability and access latency, power efficiency, however, is not explored.
	 
	Works have also been done towards preserving the data present in the chips in order to not incur an additional re-computation cost. Prior works on saving the state of the data on caches are done in~\cite{leecho,ozturk,wang}. In~\cite{wang}, the authors propose a technique to design a reliable SNUCA cache for an NoC based CMP. The authors first demonstrate how the chip will degrade over time and subsequently remapped the data without any loss.

\begin{table}[]
\centering
\resizebox{\textwidth}{!}{%
\begin{tabular}{|l|l|l|l|l|l|l|}
\hline
\multirow{2}{*}{\textbf{Sl. No.}} & \multirow{2}{*}{\textbf{Name}} & \multirow{2}{*}{\textbf{Savings}} & \multicolumn{3}{l|}{\hspace{2cm} \textbf{Overhead}}                                   & \multirow{2}{*}{\textbf{State}} \\ \cline{4-6}
   &                   &         & \textbf{Performance} & \textbf{Area}                        & \textbf{Power}   &           \\ \hline \hline
1. & Rank Aware~\cite{lu2016}        & 24\%    & 4\%   & 0.4\%                       & 1.40\%  & Cached    \\ \hline
2. & PRA~\cite{lee2017}               & 23\%    & -     & 3\% of a 2 Gb DRAM          & 0.017\% & N/A       \\ \hline
3. & RAIDR~\cite{raidr}             & 20\%    & -     & 0.013 mm\textsuperscript{2} & -       & N/A       \\ \hline
4. & SPBR~\cite{isqed}                  & 9.11\%                   & 0.82\% & \textless 1\% of 400mm\textsuperscript{2} die & 1.09\% & Preserved              \\ \hline
5. & Compiler Directed~\cite{delaluz-bank} & 23\%    & -     & -                           & -         & N/A       \\ \hline
6. & Voltron~\cite{reduced_volt}           &  10.5\%       & -       &  -                           &  -       & N/A       \\ \hline
7. & BAMM~\cite{10.1109}           &   6.30\%      &  0.70\%     &  -                           & -        & N/A       \\ \hline
8. & TAP-low~\cite{liu2010} & 19.90\% &	7.70\% & - & - & N/A \\ \hline
9. & \textbf{VAR-DRAM}          & 29.54\% & 0.8\% & 0.002\%                     & 1.09\%  & Preserved \\ \hline \hline
\end{tabular}%
}
			\caption{Summary of Proposed and Existing Works}
			\label{tab:summary}
\end{table}		

	To regulate the supply voltage on a device, researchers widely use the concept of power gating~\cite{pg}. This reduces the leakage current of the chip 
	which makes power gated \textit{static random access memories} (SRAM) circuits for implementing sleep mode, which is considered an efficient power management technique. In~\cite{srampg}, the authors propose a \textit{quasi-power-gating} approach in order to reduce leakage power dissipation on SRAM banks. Other applications of power gating are observed in \textit{network-on-chip(s)} (NoCs)~\cite{noc1} and Non-Volatile (NV) memory technologies~\cite{nvpg} where the authors proposed a power gated 1 Mb NV embedded memory to optimize the trade-off between macro size and its operational power.
	
	\subsection{Limitations of Existing Works}
	    We agree that most of the aforementioned prior works promise significant savings in terms of DRAM energy. Table~\ref{tab:summary} showcases most of the previously discussed works highlighting DRAM power savings in contrast to its respective additional hardware requirements. Due to the conventional structure of DRAM devices, most DRAM power or energy-saving works are directed towards a \textit{rank-aware} power management mechanism~\cite{ddr4,lu2016,delaluz,lebeck,raidr}. Only a handful of works are directed towards saving power from DRAM banks~\cite{delaluz-bank,isqed}. Compiler-directed power management~\cite{delaluz-bank} may not be always feasible as it'll limit the distribution of pre-compiled files. Power management mechanism at a finer granularity than ranks is becoming a necessity for memory devices as the demand for low-powered electronics and mobile devices are escalated over the last decade. DRAM manufacturers have partially addressed this issue by introducing per-bank refreshing in low-powered DDR devices. However, there seems to be further room for improvement regarding this respect. The DRAM device's capacity may not be completely filled constantly. However, both static and dynamic power in terms of background and refreshes would be consumed by these devices. This constitutes our first observation from prior works. 
	    The concept of trading off memory capacity for lower latency is not new. Choi et al.~\cite{mcrdram} propose a mechanism to clone multiple rows which in turn allows faster accesses. This is due to the fact that the speed of the sensing process increases as the number of sensed-cells is more. Luo et al. propose CLR-DRAM~\cite{clrdram}, where a similar concept is used as the authors use more than one sense amplifier to allow faster accesses of cloned rows. 

	    

	    The deviation from nominal latency and power consumption due to process variation in modern-day DRAM devices calls for utmost attention. Process variation is a well-documented phenomenon for DRAM devices~\cite{process,o34,vampire}. Yet, most of the aforementioned works fail to accommodate a technique to handle process variations. It has been predicted that in the future, the effects of process variations are likely to increase~\cite{sarangi}. In this context, the under-performing cells of the DRAM device should either be mitigated by remapping or, turned off. The current memory technology does allow remapping~\cite{sample} but doesn't allow turning off due to the inherent design of the entire memory circuit. It becomes logical only if a mechanism exists which can extract power savings from these unused cells at a cheap cost. Modest modifications to the DRAM circuitry by using power gating may allow the aforementioned problem to be rectified. Power gating is a technique that has found significant use in SRAM and NoC architectures, to limit leakage current in SRAM technology but has been sparingly explored in the context of DRAM. Again, scaling becomes an issue because a modern-day DRAM device has 32,768 rows, and power gating every single row is infeasible. DRAM banks, therefore, would become the ideal selection for power gating in this context, which will solve the previously mentioned issue of mechanisms lacking \textit{bank-wise} power management for DRAM devices. Moreover, such a technique can be independently placed into the memory device, which, would further allow other power-saving mechanisms to be incorporated. As a consequence of avoiding variation affected locations, the latency of the DRAM device ameliorated, closing the gap toward an ideal device. Moreover, while powering down, we also include a search and storage efficient \textit{light-weight} translation mechanism. This allows preserving the state of the DRAM device during such instances by redirecting variation affected addresses to their new location. 
	    
	    
	    


%% file: Text/background.tex
\section{Background}
\label{sec:background}

This section is divided into three subsections, each explaining a concept used in VAR-DRAM. 

\input{Text/dram_background}

\subsection{Process Variation (PV)}
    \label{sec:provar}
         Process Variation is defined as the deviation of process, voltage, and temperature from the nominal specifications. As per the traditional Shockley model, we see that the  effective channel length ($L_{eff}$) and threshold voltage ($V_{th}$) are the primary factors while characterizing CMOS delay under process variation. PV is characterized by the following:
         
        \hspace{-0.425cm}    1. $V_{th}$ is a driving factor in process variation. Transistor drain current ($I_d$) is directly dependent on $V_{th}$ which is given by~\cite{kang}:
            
        \begin{equation}
        I_d = 
        \begin{cases}
            \text{$0$,} &\quad\text{if $V_{gs} \leq V_{th}$}\\
            \text{$\beta(V_{gs} - V_{th} - \frac{V{ds}}{2})V_{ds}$,} &\text{if $V_{ds} < V_{gs} - V_{th}$}\\
            \text{$\beta \frac{(V_{gs} - V_{th})^2}{2}$} &\text{if $V_{ds} \geq V_{gs} - V_{th}$}
        \end{cases}
        \end{equation}
        Here, $\beta = \mu C_{ox}W/L_{eff}$, where $\mu$ is mobility and $C_{ox}$ is oxide capacitance.
        
        \hspace{-0.425cm}    2. Switching time of the circuit ($T_g$) is given by
        \begin{equation}
        T_g \propto \dfrac{L_{eff}V}{\mu(V - V_{th})^2}
        \end{equation}
        Hence, we also infer that PV affects the timing parameters of a chip.
        PV has die-to-die (D2D) and Within Die (WID) components. WID can further be divided into random (rand) and systematic (sys) sub-components.  In collective form, we can write:
        \begin{equation}
        \Delta P = \Delta P_{D2D} + \Delta P_{WID} = \Delta P_{D2D} + \Delta P_{rand} + \Delta P_{sys}
        \end{equation}

        
    \subsubsection{Systematic Variation}
        \label{sec:sysprovar}
        Systematic Variation, which is prevalent in DRAM systems, exhibits spatial correlation. This implies that nearby transistors also share similar systematic parameter values~\cite{sarangi}. It follows a multivariate normal distribution. It is assumed that for a pair  of points $m$ and $n$ on a chip, the spatial correlation is only dependent on the distance between $m$ and $n$~\cite{sarangi}. The correlation function is given by:
        \begin{equation}
            corr(P_x, P_y) = \rho(d) , \quad d = |x-y|
        \end{equation}
        If $d = 0$, then $\rho(d) = 1$. This means that \textit{it is completely correlated.} The spherical function $\rho(d)$ is defined as

     \begin{equation}
    \rho(d) = 
     \begin{cases}
       \text{$1 - 3d/2\phi + d^3/2\phi^3$,} &\quad\text{if $d \leq \phi$}\\
       \text{$0$,} &\quad\text{otherwise}
     \end{cases}
        \end{equation}
            
        The parameters of mean ($\mu$), variance ($\sigma^{2}$), and density ($\phi$) are provided according to the technology node used~\cite{technode}, which we have incorporated while implementing the proposed technique. The value of $\sigma$ is calculated from the ratio of $\mu/\sigma$. The ratio differs for different technology nodes. $\phi$ varies accordingly.  

        
    \subsubsection{Access Latency}
        \label{sec:acclat}
        Access latency is defined as the time required for an instruction to command a transition and the underlying hardware to actually execute the same by transitioning the voltage to either high or low. PV affected cells of the DRAM devices exhibit higher latency. Prior works have stated that it is essential to determine the \textit{critical path} of a DRAM device before studying the effects of variations. Studies like~\cite{process,ras_timing} have concluded that there is nearly an increment in accessing the row decoder by 48\%. The latter provided an absolute increment value for tRAS as 18 ns. Naturally, the longer the time is taken to access data in the DRAM device, the higher the energy consumption. Healthy banks show the least access latency. This fact is used to calculate the effective access latency of all other banks via comparison of these specific banks. 
        
        \begin{figure*}[tb]
        \begin{center}
            \mbox{ 
            \hspace{-1.0ex}
            \subfigure[VARIUS~\cite{sarangi} generated map for $L_{eff}$]
                {
                \label{fig:leff_map}
                \includegraphics[width=0.48\textwidth]{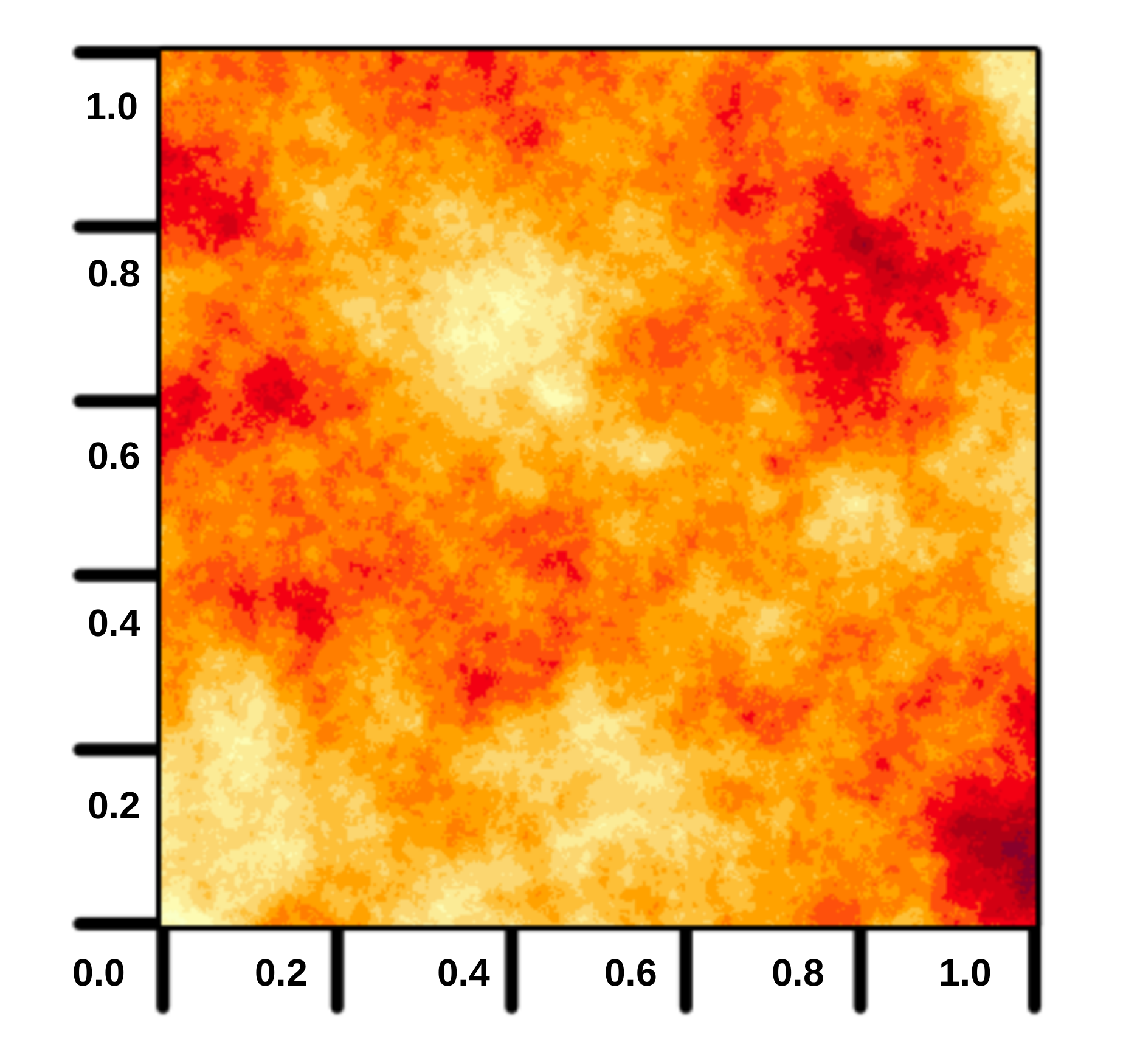}
                }
            \hspace{-2.0ex}
            \subfigure[Diagram of a Simple 4 Level Trie]
                {
                \label{fig:trie}
                \includegraphics[width=0.48\textwidth]{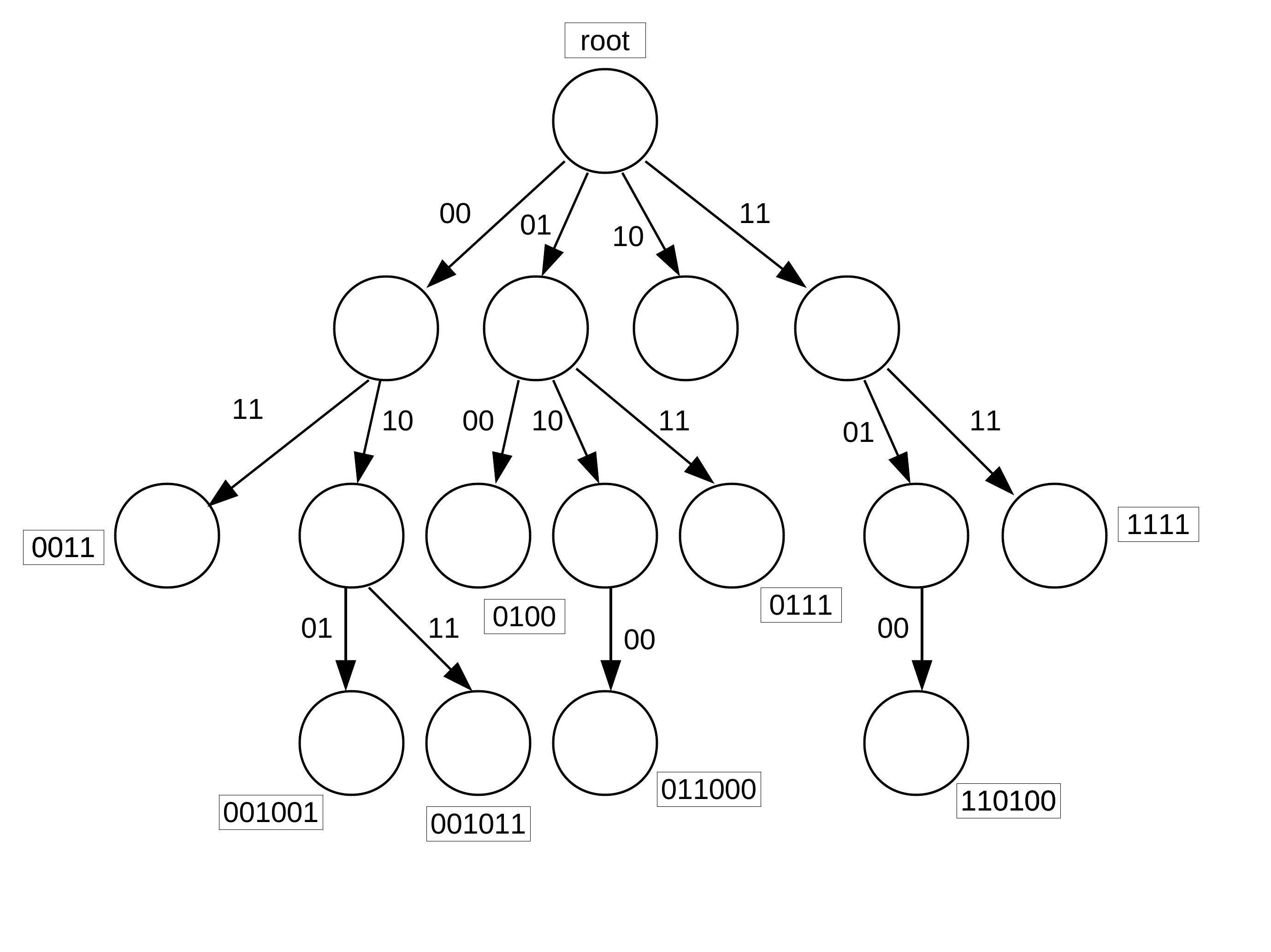}
                }
            \hspace{-2.0ex}
            }
            \caption{Variation Distribution Map (left) and a simple Trie Structure (right)}
            \label{fig:variusandtrie} 
        \end{center}
    \end{figure*}
    
    \subsection{Retention Time}
        \label{subsec:retention}
        The refresh window (tREFW) of a modern-day DRAM is set at 64 ms in normal temperature and 32 ms in case of higher operating temperature ($\geq 85^{\circ}C$)~\cite{ddr4}. This defines the worst-case value for the entire DRAM chip. However, not all DRAM cells require recharging at 64 ms~\cite{crow,raidr} as there only exist a few rows which have retention time less than equal to 64 ms. There has been several research works previously done in order to identify such weak rows~\cite{liu-retention,retention-paper-1}. Most of these profiling techniques would pass a stream of data patterns within a wide range of operating temperatures to pinpoint DRAM cells failing to meet a set retention time threshold. Liu et. al~\cite{liu-retention} concluded that there exist approximately 1000 weak rows in a 32 GB DRAM device. Yet, even for this small fraction of rows, a DDR4 DRAM device is uniformly refreshed at 64 ms in normal working temperature. Hassan et. al.~\cite{crow} calculated that the bit error rate (BER) of the 4X refresh window DRAM device (tREFW = 256 ms) is $4 \times 10^{-9}$. Further, the authors gave the probability of the presence of a weak cell in a row as:
        \begin{equation}
            \label{eq:ret}
            P_{weak\_row} = 1 - (1 - BER)^{N_{cells\_per\_row}}
       \end{equation}
       Here, $N_{cells\_per\_row}$ is the number of weak cells per row. Clearly, the number of weak rows present in a DRAM bank would be very small. 
        
        \subsubsection{Variable Retention Time}
            \label{sec:vrt}
            Variable Retention Time, or simply \textit{VRT}, is a recent phenomenon observed due to variations~\cite{avatar,approx2}. It is a subset of the previously discussed retention time. The major distinction is the fact that VRT affected DRAM devices exhibit different retention times during runtime. A given row may be mapped to a retention time window of 128 ms. However, due to the effect of VRT, present on newer DRAM devices, the same row may vary its retention time to more than 128 ms or even less than 64 ms. Qureshi \textit{et al.}~\cite{avatar} have shown that approximately 31,798 unique rows are affected by VRT in a DRAM rank over a period of 30 days. This renders conventional retention-time profiling techniques, where a given row is mapped to a specific retention time, ineffective for modern DRAM devices. In Section~\ref{sec:related}, we have discussed about approximate DRAM devices. Such techniques exploit error-tolerant applications against variations.
        

%% file: Text/dram_background.tex
\subsection{Dynamic Random Access Memory}
    This subsection is further divided explaining the organization, working, and, core power components of a typical DRAM device.
    \subsubsection{Organization and Working}
    \label{DRAM-operation}
    A modern-day DDRx DRAM device is organized in six hierarchical levels. The memory channel sits at the top which is operated concurrently with other channels. A DRAM can accommodate multiple Dual Inline Memory Modules (DIMMs), which is a module containing one or more random access memory (RAM) chips. These are divided into ranks, which are used to distinguish DIMM level independence and internal bank-level independence. The rank is further divided into chips and then into independent banks that provide the lowest level of independent operation. Each bank has its own row decoder. DRAM banks are further divided into rows and then columns. 
        
    The master operation  of the DRAM device is controlled by the clock enable (CKE)~\cite{ddr4} which must be high in order for the DRAM to receive commands. The incoming command or address is pushed into the decoding logic of the DRAM. The first command sent to the DRAM is usually an Activate (ACT) command which is responsible for selecting the appropriate bank and row address. The data stored in the corresponding DRAM cells are then transferred to the sense amplifiers which retain the data until a Precharge (PRE) command to the same bank is issued. The equivalent time required is called row cycle time (tRC), which can be written as \textit{tRC = tRAS + tRP}. \textit{tRAS} is called \textit{row access strobe}, which is the time interval between row access command and data restoration in a DRAM array. \textit{tRP} is called row precharge time, which is the time interval that it takes for a DRAM array to be precharged for another row access. Every ACT command has to have a PRE command associated with it. A READ or a WRITE can only be performed by the DRAM in its active state. 
        
    \subsubsection{Power Components and Power Management}
    \label{subsec:dram-pow}
    Power in a typical DRAM is consumed under five components: Background, Activate; Precharge, Read/Write, On-Die Termination (ODT), and Refresh. CKE dictates the amount of current to be consumed in a DRAM device. In the active state, a higher current is consumed by the device. If there is a prolonged period of inactivity of the device, CKE is set to low which consumes a lower amount of current. This management scheme is implemented in a per rank manner in all existing DRAM devices and is usually invoked whenever the command queue of the DRAM device is empty for a longer period of time~\cite{dramsim,micro,ddr4}. It is the most popular power management technique employed in DRAMs. However, irrespective of operations, the device always consumes a static power which is called the background power of a DRAM device. The fraction of power consumed under any ACTIVATE (ACT) and/or PRECHARGE (PRE) command constitutes Activation; Precharge power. A Read or Write consumes power depending upon the fraction of time the data was on the bus. The system-dependent power components, like output driver power, constitute ODT power. Since the DRAM is a capacitor-based technology, it requires periodic recharging of the capacitors in order to retain the data it holds. The power consumed for the aforementioned periodic recharging of the DRAM capacitors, constitutes the refresh power.

%% file: Text/methodv2.tex
\section{Methodology}
\label{sec:method}

    In this section, we discuss our proposed technique. VAR-DRAM first requires a list of variation affected banks in order to initiate \textit{migrate and remapping} policy. Initialization phase of VAR-DRAM (\S~\ref{subsec:init_me}) is either statically programmed by the DRAM manufacturer with variation-affected banks or dynamically change with variation sensor data. Data remapping is a costly operation in context of DRAM devices~\cite{rowclone}. The remapping phase (\S~\ref{subsec:mig_over}) is divided into several sub-sections (\S~\ref{subsubsec:trie}, \S~\ref{subsuec:mig}, \S~\ref{sec:data}, and, \S~\ref{sec:collision}). There are two sub-phases of remapping: (a) address remapping  (\S~\ref{subsuec:mig}), and, (b) data remapping (\S~\ref{sec:data}). We propose a logarithmic look-up times for migrated addresses  (\S~\ref{subsubsec:trie}). Edge cases are discussed Section~\ref{sec:collision}. The key design elements of VAR-DRAM include a variation-aware memory controller with a modified power management unit, a modified address mapping policy, and, power gated DRAM banks. The working of the proposed technique is explained in Section~\ref{assembly}. It is followed by the hardware implications (\S~\ref{sec:hw_impl_para}). 
    As a conclusion of this section (\S~\ref{subsec:benefits}), we explain how VAR-DRAM is beneficial for DRAM devices.

    \subsection{Initialization Phase}
        \label{subsec:init_me}
        As mentioned before, VAR-DRAM requires a list of variation affected banks to begin with. A two-dimensional matrix, called \textit{variation matrix (V)}, is used for the aforementioned task. This matrix will either be pre-programmed with variation-affected cells provided by a DRAM manufacturer~\cite{sample}, or, dynamically updated via the usage of variation detection sensors which will be installed in the DRAM array~\cite{vsensors}. In either of these cases, \textit{V} is updated alongside normal DRAM operations. During simulations, we have used a statistical PV model, called VARIUS model~\cite{sarangi} to identify variation affected banks present in the DRAM device. A variation map is generated and is divided into \textit{number of ranks} $\times$ \textit{number of banks}, corresponding to a chip floor plan. The direct effects of variation in terms of timing is previously explained in Section~\ref{sec:provar}, Section~\ref{sec:sysprovar} and Section~\ref{sec:acclat}. These are incorporated while simulating the variation map during simulations. The variation map signifies the distribution of $L_{eff}$. Naturally, variation affected banks are slower. We define \textit{victim} and \textit{target} bank pairs henceforth, which, signifies a variation affected bank (slow) and a healthy bank (normal) respectively. \textit{V} is programmed accordingly. A victim bank shows characteristics of a variation affected bank, such as an increased access latency of the bank. These banks are programmed with higher tRAS value as reported in previous works including~\cite{process,ras_timing}. The timing of healthy banks follow an ideal DRAM device.
        
    \subsection{Remapping Phase}
        \label{subsec:mig_over}
        Our aim is to extract power savings from victim banks. We achieve this by powering down these banks. But addresses and their respective data contents have to be preserved in order to avoid data corruption during execution. Address migration depends on the method of operation used during execution, whether, banks are powered down at the start of the system (\textit{static closing}) or are powered down during execution (\textit{dynamic closing}). In the former case, we need to translate a given victim address to their corresponding target address. In the latter case however, we would also require to migrate data from these addresses to its corresponding target addresses. We use a table of remapped addresses. The table is implemented using a data structure called \textit{trie} as it provides space efficiency over simple tables with the trade-off with logarithmic lookup times. In this subsection, we further explain the aforementioned components in details.
        
        \subsubsection{Trie Implementation}
            \label{subsubsec:trie}
            By definition, a trie is a type of search tree, which is an ordered tree data structure. It is used to store a dynamic set or associative array where the keys are usually strings. It is mostly analogous to a binary search tree but unlike a binary search tree, no node in a trie stores the key associated with that node. Theoretically, it gives search time as $O(l)$, where $l$ is the length of an address and has a better space complexity than a table. It is feasible to be implemented in Silicon~\cite{trie_paper}. Hence, it becomes our ideal candidate for storing a table in the lowest space requirements. We have addresses incoming to the memory, which requires a dynamically allocated space structure while maximizing the number of possible stores. A node's position in the trie defines the key with which it is associated. Each node in our implementation has 8-bit values giving a total of 4 levels. In order to further reduce the number of cycles required to traverse the trie, we use dual-edged flip flops~\cite{dualedgeFF}. Both the rising and falling edge of the clock are then effectively used for transitioning between nodes of a trie. This reduces lookup time by half.


        \subsubsection{Address Remapping}
            \label{subsuec:mig}
            During address remapping, we operate on a simple format, \textit{i.e.}, rank and bank pair. These are changed in case of a target address, while keeping row and column numbers the same in the target address. \textit{V} already provides target rank and bank for a given pair of victim rank and bank. The modified address mapping policy uses these numbers to get final or effective values of rank and bank. However, there may be instances of addresses being collided with each other, which we discuss later in this section. During static closing, victim addresses are simply translated. However, for dynamic closing, addresses belonging to victim locations need to be stored and migrated first. For storing this set of addresses, we propose a small storage space alongside the trie-based table. We discuss its absolute size required/used during evaluation in Section~\ref{sec:result}.
            
            Similar to the powering down of banks, we use the same mechanism for switching previously closed banks. This mechanism is particularly necessary for memory-intensive benchmarks. A memory device with powered down banks will show direct benefits in terms of energy savings. However, it also implies that the system is essentially operating at a smaller memory size than its original counterpart. We therefore preemptively power on previously powered down the bank. This is done so that the memory does not reach a throttling point, where memory capacity runs out of working memory set size.
		
		    Algorithm~\ref{alg:migrate} gives an overview of how the modified memory controller functions. While static closing is invoked, there is no requirement of data remapping, and the proposed method is limited to translations only. Address translation is done in a trivial manner initially in \textit{line 7} of Algorithm~\ref{alg:migrate}. In parallel, once the victim rank and bank pair is identified, it is passed for collision checking, whose functionality is discussed in Section~\ref{sec:collision}. In a simple translation, a pair of target rank and bank pair is returned, which is computed in parallel while decoding an address. Fetching a rank and bank pair requires 16 to 24 bits: 15 bits for a DRAM bank and rest 7 bits for accommodating and scaling variable number of DRAM ranks. Once, the target bits are fetched, 2 additional shift operations yield the final target rank and bank bits. The memory controller receives the effective address and then proceeds normally. However, if an INTERRUPT is raised due to address collision, the previous memory access is invalidated, stalling 3 clock cycles, which is required to compute the new address with updated row and column values. Collisions happen rarely. As a consequence of the same, the overall impact of such stall cycles is limited. However, uniformity of timing has to be maintained in DRAM devices as these devices are strictly based on timing~\cite{ddr4,micro}. We include this lookup time within tRAS time for all victim banks. The memory controller is patched with two tRAS values corresponding to memory accesses made to the victim and target banks' addresses. VAR-DRAM therefore never violates any timing constraints of a conventional DRAM device.
		    
		    Powering down banks has two phases. A flag bit(s) of 3 values is used. It is set at \textit{00}, if switching is not required. If there is a remapping call, the flag is set to \textit{01}, which indicates migration has started. After data in all victim addresses are remapped, the flag is set to \textit{10} indicating the power management unit to power down victim banks. While powering up, the same is set from \textit{10} to \textit{00}, as when all banks are operational, there is no need for address translation. The MSB of the FLAG bit is used for the select line in case of both the DEMUX and MUX units. 
		    
                \begin{algorithm}
			\caption{Address Remapping Scheme}
			\label{alg:migrate}
			\begin{algorithmic}[1]
			\STATE // V is an M $\times$ N array representing the \textit{variation matrix}
			\STATE Ensure V is valid
			\STATE // A flag FLAG  ($00$) represents reconfigured banks
				\WHILE {valid $address$ is incoming}
				\STATE // Resolves address into its corresponding \textbf{ch}annel, \textbf{ra}nk, \textbf{ba}nk, \textbf{ro}w and \textbf{co}lumn
				\STATE ch, ra, ba, ro, co = resolve($address$, FLAG)
				\STATE // Check for address collision in parallel. Sends INTERRUPT if ro, co mismatches
				\STATE ro', co' = collision\_check ($\tau$, address)
    		    \IF {INTERRUPT}
    				    \STATE raise INTERRUPT, use ch, ra, ba, ro', co'
				\ENDIF
				\STATE update\_power (FLAG)
				\STATE //Continue normal memory operation
				\ENDWHILE
			\end{algorithmic}
		    \end{algorithm}		    
		    
        \subsubsection{Data Remapping}
        \label{sec:data}
            In VAR-DRAM, data remapping or migration is invoked in two cases: (a) dynamic closing, and (b) banks reopening. Based on a rearranged trie, a migration of either kind is issued. This possesses an overhead on the memory bandwidth. There are several works done previously in order to initiate bulk copying in DRAM devices~\cite{intelpaper,rowclone}. For our proposed work, we have used Rowclone~\cite{rowclone}. For migrations, we copy the contents of an address (column width) from one bank to another using RowClone's PSM mode. Latency and bandwidth of data migration carried over from RowClone. We patch the memory controller with data remapping capabilities and record its outcome. Overhead in terms of latency, as well as bandwidth during data remapping, are incurred. Furthermore, VAR-DRAM executes programs with a reduced latency device. In most cases, data remapping latency is compensated via the usage of faster memory locations.
            
            We perform bulk remapping based on Algorithm~\ref{alg:migrate2}. We require the number of victim bank and target bank count to be equal for pair formation. \textit{srcBank} and \textit{destBank} are two lists whose relation is maintained via the index (\textit{i}). For each bank in \textit{i}\textsuperscript{th} index, we recursively obtain all victim addresses via traversing the trie $\tau$. The address is then passed to \textit{storeContent} with parameters \textit{srcAddr} and \textit{destBank}. Remapping takes time. The inconsistency is likely to arise is when there will be a write access to a victim address during migration. We, therefore, maintain a small table in the proposed remapping unit, where write access' addresses are put in the format \textit{addr, flag\_bit}. The working of the remapping unit is later discussed in Section~\ref{assembly}. Inside \textit{storeContent}, this table is first checked. If there is the \textit{flag\_bit} set, the remapping of this row gets a higher priority. The write access is stalled (at most by \textit{RowClone's write latency + table search time}) and is served then. When this address appears for its scheduled remapping in Algorithm~\ref{alg:migrate2}, the request is ignored. \textit{storeContent} also checks for address collision, which is discussed in Section~\ref{sec:collision}. Data remapping is performed in parallel to normal DRAM operations. Bandwidth is used alongside normal operations. In Section~\ref{sec:result}, we show that this has no direct effect on normal DRAM operations.
            
        \subsubsection{Address Collision}
            \label{sec:collision}
            Issuing a new address for a victim bank to a target bank introduces a probability of collision with an existing address corresponding to the target bank. Our model handles this in a na\"ive but efficient manner. If a collision occurs, the column address in the modified address mapping policy is incremented until the next immediate free space is found. This would mean that a simple translation of rank and bank will not be sufficient. Here, the role of trie becomes significant. The trie is iterated for all incoming addresses. However, this is done in parallel with the normal DRAM functioning and starts whenever address decoding begins. Collectively, a full iteration of the trie for the victim address is completed 3 clock cycles, and, retrieval takes another 3 clock cycles, which lies outside of the critical path by 3 clock cycles. An INTERRUPT signal is sent to the memory controller nullifying the previous memory transaction and new addresses are used to further continue the memory transaction, which is represented in line 9 of Algorithm~\ref{alg:migrate}.

            \begin{algorithm}
            \caption{migrateAndRemap($\tau$, srcBank{[]}, destBank{[]})} 
            \label{alg:migrate2}
            \begin{algorithmic}[1]
            \ENSURE FLAG $\in {00, 01, 10}$ 
            \REQUIRE size (srcBank[]) == size (destBank[])
            \STATE FLAG = $01$ // or $10$
            \FOR{$i \leftarrow size (srcBank[])$}
                \STATE srcAddr[] = $\tau$.recursiveGet (srcBank[i])
                \FOR{$j \leftarrow size (srcAddr[])$}
                    \STATE storeContent(srcAddr[j], destBank[i])
                \ENDFOR
            \ENDFOR
            \STATE FLAG = $10$ // or $00$
            \STATE return $\tau$
            \end{algorithmic}
            \end{algorithm}
        \subsubsection{Target Bank Overflow}
            \label{sec:overflow}
            The overflow of space for a target bank is a boundary case for VAR-DRAM. We are maintaining a trie consisting of all unique addresses. Hence, we can find the number of active addresses for a particular DRAM bank. If this number exceeds 90\% of the capacity of the bank, we simply initiate a reverse migrate signal to the corresponding victim and continue the execution of the program. During evaluation in Section~\ref{subsec:overhead}, we further explain why this case is unreachable using configurations in this particular iteration.

    \begin{figure*}[t]
        \begin{center}
            \mbox{
            \hspace{-1.0ex}
            \subfigure
                {
                \includegraphics[width=0.98\textwidth]{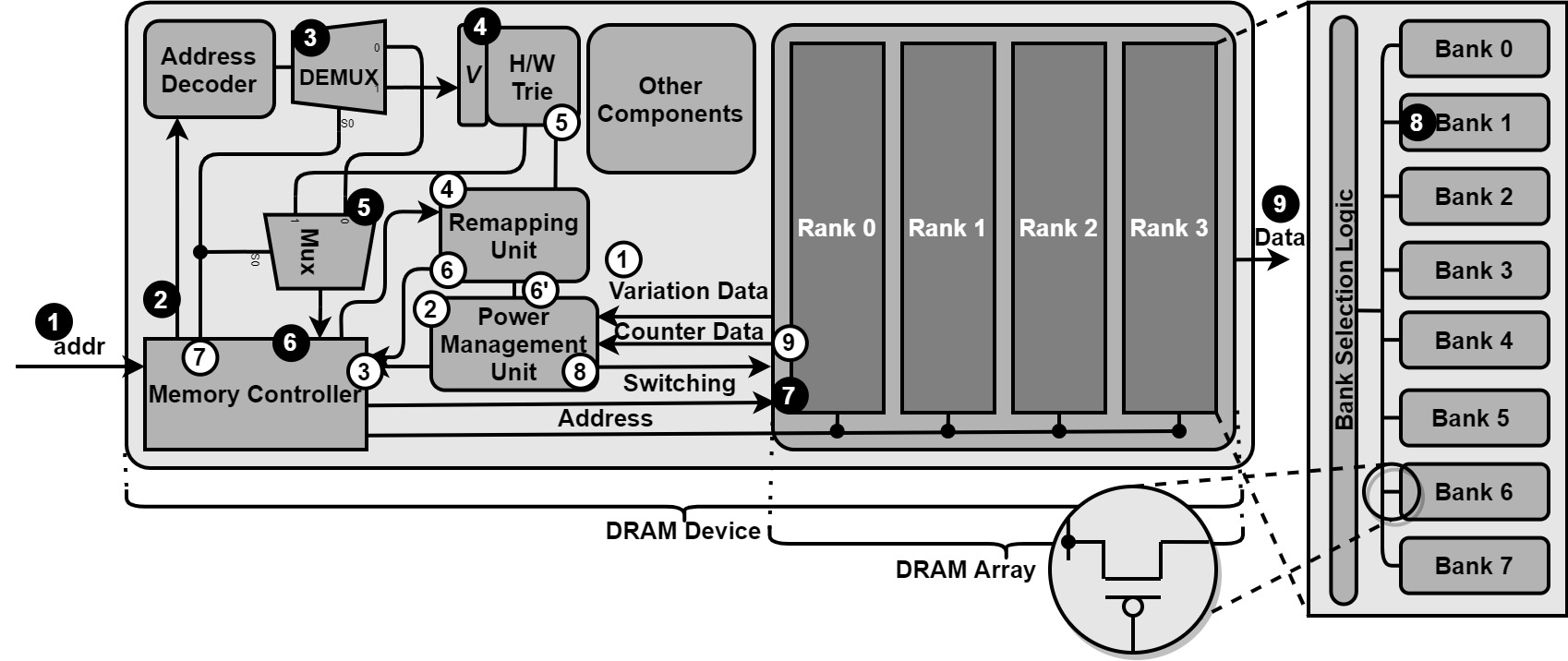} 
                }
            }
            \caption{Functional block diagram of the proposed mechanism. The DRAM device has 4 ranks, each containing  1 chip. Each chip further contains 8 power-gated banks. There are 2 events shown in the diagram: (a) address translation, and, (b) bank-level powering down.}
            \label{fig:func} 
        \end{center}
    \end{figure*} 

    \subsection{Assembling VAR-DRAM}
        \label{assembly}
        In this subsection, we discuss the collective working of VAR-DRAM. The working of each individual components is explained previously. Figure~\ref{fig:func} represents the functional diagram of the proposed technique. The existing component in a conventional DRAM sub-system are the memory controller with an address decoding logic, power management unit, and the DRAM array. We propose the addition of a trie and remapping units alongside the modification to the power management unit. There are two events represented in the figure. The one followed by \encirclewhite{\textit{i}} represents the dynamic switching of DRAM banks. Address redirection in the case of a previously powered down bank is represented by \encircleblack{\textit{j}}. 
        
        We first explain the switching mechanism of VAR-DRAM. In Section~\ref{subsec:init_me}, we have explained the fact that variation sensors report deviations from the nominal access latency of the DRAM device. The aforementioned variation data is shared by these sensors to the power management unit (\encirclewhite{1}), which contains instructions for enabling power gating of DRAM banks. A list of the victim and target bank pairs (V) is then generated in the \textit{power management unit} (\encirclewhite{2}). The list is then shared with the \textit{memory controller} (\encirclewhite{3}) by the power management unit. If the model follows dynamic switching, then the remapping phase begins (Section~\ref{subsec:mig_over}) at this step. In Action~\encirclewhite{4}, the memory controller sends the list of victim and target pairs to the \textit{remapping unit}. The remapping unit then proceeds to rearrange data in the trie. Addresses are rearranged within the trie in Action~\encirclewhite{5}. Data remapping then is initialized, which follows Algorithm~\ref{alg:migrate2}, and the FLAG bit is set to \textit{01}. The working of the FLAG bit is previously discussed in Section~\ref{subsuec:mig}. The FLAG bit value is shared with the memory controller. During this phase, the address translation is inactive. Once the FLAG bit changes to \textit{10}, an \textit{acknowledgment} is sent back to (a) the memory controller (\encirclewhite{6}), and, (b) the power management unit (\encirclewhite{6'}). This change in the value of the FLAG bit acts as this aforementioned acknowledgment. This effectively enables the VAR-DRAM logic sequence. Address translation (\encirclewhite{7}) is then in effect until the value of the FLAG bit changes. The select line for the DEMUX and the MUX is the MSB of the FLAG bit, which was previously explained in Section~\ref{subsuec:mig}. In case of \textit{static closing}, this data is shared at startup, which skips Actions~\encirclewhite{4}, \encirclewhite{5}, \encirclewhite{6}, and, \encirclewhite{6'}. Action~\encirclewhite{8} activates the switching mechanism, powering down the victim banks. Counter data, signifying utilization, is then monitored by the power management unit (\encirclewhite{9}) for powering up the victim banks. The counter data acts as a preemptive indicator, that the capacity of the DRAM device is running out. Counter data is periodically generated from the size of the effective size of the trie. Whenever counters report a specific percentage of the total memory usage, powering up action(s) comes into effect. This value is experimentally found out. It is discussed in Section~\ref{sto_ov_x} in details. Powering up follows a similar mechanism with the exception of the FLAG bit value, which changes from \textit{10} (powered down state) to \textit{00} (no translation required state).
        
        Whenever an address appears into the memory system (\encircleblack{1}) for I/O operation in the memory controller, its effective address is computed by the modified \textit{address decoder} (\encircleblack{2}). The modification of the same largely includes the proposed trie unit. We have installed a DEMUX unit (\encircleblack{3}) to select the mode of operation, in case VAR-DRAM is active. The select line for the DEMUX unit is guided by the MSB of the FLAG bit. If VAR-DRAM is active, then the address' corresponding target rank and the bank is extracted from the H/W trie (\encircleblack{4}). This was previously explained in Section~\ref{subsuec:mig} and Algorithm~\ref{alg:migrate} (Lines 5 -- 8). The final address is selected using a MUX unit (\encircleblack{5}). The select line for the MUX unit is also the MSB of the FLAG bit. The final address is then forwarded to the memory controller (\encircleblack{6}). The aforementioned actions are collectively depicted in Algorithm~\ref{alg:migrate}. The decoded bits are then sent to the DRAM array (\encircleblack{7}) for data retrieval. The data is finally obtained from its corresponding DRAM rank, bank, effective row, and effective column. This is represented in Action~\encircleblack{8}. Finally, the data is sent to the memory bus (\encircleblack{9}). 

    \subsection{Hardware Level Implementation}
        \label{sec:hw_impl_para}
        The power gated design for the proposed DRAM chip is described in Figure~\ref{fig:func}. We have attached sleep transistors for power gating on each bank of a DRAM rank. Also, each of these banks has its own utilization counter to report excessive usage. The utilization compute unit reports the DRAM bank's utilization to the control unit which also receives information on decaying portions of the DRAM chips. We proceed with the assumption that fabricated end processors will have variation sensors~\cite{vsensors} present on the chip in order to collect statistics on decaying portions of the chips. These variation sensing sensors are placed on each bank of the DRAM chip and will report their respective findings to a control unit.
    
        The control unit will forward a decision to the Power Management and Re-Mapping Unit to decide whether a DRAM bank has to be explicitly powered down or not. If performance is degrading at a faster rate than decaying chips, we do not attempt power down. Also if the memory requirements of the end application increase, we turn on powered down banks preemptively. During our experiments, we have set a threshold in terms of the storage area of the trie as an indicator of degradation. This is further discussed in Section~\ref{sto_ov_x}.
    
        The Power Management and Re-Mapping unit will invoke the power gating module which is implemented in a conventional manner with a sleep transistor inserted between the supply voltage (V$_{DD}$) and the bank selection logic. This unit will also handle data migration from victim banks to target banks.  

    \subsection{Implications of VAR-DRAM}
        \label{subsec:benefits}
            The benefits of VAR-DRAM can be unfolded into four distinct categories: (a) efficient energy usage, (b) lower memory latency, (c) reduction of refresh instructions, and, (d) higher reliability of data. We explain these in this sub-section.
            
            \subsubsection{Energy Usage}
                \label{subsec:m_energy}
                Variation affected DRAM devices require higher operating energy as these devices exhibit higher latency. Completion of one memory access requires more time, and, energy. VAR-DRAM eliminates all such instances. Moreover, the powering down of banks using power gating limits the amount of leakage power associated with these banks. This translates to a lower energy consumption of the device. The memory device is never on its bottleneck point as we power up previously closed banks preemptively. The full potential of VAR-DRAM can be exploited with an OS-transparent memory policy. If the remapping logic is provided with a list of invalid pages, banks can be dynamically powered on and off whenever required. Nevertheless, in all cases demonstrated in this paper, we gain a positive energy savings value.
                
            \subsubsection{Reduced Memory Access Latency}
                \label{subsec:m_latency}
                As we avoid mapping addresses to variation affected areas of the DRAM device, there exists a lesser deviation of timing parameters from its nominal counterparts. In other words, memory accesses are close to ideally operating devices. Although we save time in this regard, however, there exists a wider impact of powering down banks in this case. In Section~\ref{DRAM-operation}, we have explained the presence of one row decoder per bank. Powering down banks reduces the number of such row decoders. As memory patterns are irregular and workload-dependent, we either would get overall faster or slower execution times. In Section~\ref{sec:result}, we further demonstrate two cases: (a) worst-case latency times including data remapping, and, (b) execution time.
                
            \subsubsection{Reduction in Refresh Instructions}
                \label{subsec:m_retention}
                In Section~\ref{subsec:retention}, we have already explained the presence of weak rows which compels all DRAM devices to refresh at a period of 64 ms. A tREFI instruction is issued after each 7.8 $\mu$s~\cite{ddr4}. As we use an internal DRAM remapping logic in VAR-DRAM, we can use it as a tool to remap weak rows. Irrespective of whether we power down banks or not, this mechanism can operate independently. It can provide benefits for a reduced refresh overhead and blocking time, thus, allowing the DRAM device to provide energy savings and faster execution times. The translation mechanism will operate in a manner similar to address collision checking, previously discussed in Section~\ref{sec:collision}, which will, unfortunately, incur mandatory stall cycles whenever there is memory access for a weak row. But, as discussed in Section~\ref{subsec:retention}, the number of weak rows is very low, which restricts overhead to a marginal value, later shown in Section~\ref{sec:result}.
                
            \subsubsection{Higher Reliability of Data}
                This benefit is self-explanatory as there is a higher probability of data corruption in variation affected areas. This was first referred to in Section~\ref{sec:related}. As we avoid mapping data to variation affected areas, we stand with a lower probability of data corruption. Even after powering on variation affected banks, critical data can be remapped to healthy locations and the remapping logic can be used for translation. This will require either an OS-transparent paging policy or a list of severely affected DRAM rows, based on which, continued migration of such addresses can be performed.

%% file: Text/evaluation.tex


\section{Evaluation}
\label{sec:eval}
    In this section we briefly introduce the evaluation setup, benchmarks used and experiment details. 
\subsection{Experimental Setup}
    \label{subsec:exp-set}
    For simulating VAR-DRAM, we've used a modified version of a cycle-accurate DRAM simulator called DRAMSim2~\cite{dramsim}. The major modifications include the incorporation of a power-efficient memory controller which incorporates a power-efficient and variation aware module. The address mapper logic is modified and is affixed with our proposed trie mechanism. The new address mapping policy takes variations into account of \textit{original} and \textit{migrated} addresses. As VAR-DRAM is a framework, it can be configured to be used with DRAM generations including DDR3, DDR4, LPDDRx, etc. For this paper, we chose to demonstrate the capabilities of VAR-DRAM on a DDR4 DRAM device based on MICRON DDR4 technical data sheet~\cite{ddr4}.

\begin{table}[]
\centering
            \caption{Simulation Parameters}
            \label{tab:params}
\resizebox{\textwidth}{!}{%
\begin{tabular}{|ll||ll||ll|}
\hline
\textbf{CPU Parameter} &
  \textbf{Specification} &
  \textbf{Cache Parameter} &
  \textbf{Specification} &
  \textbf{Memory Parameter} &
  \textbf{Specification} \\ \hline \hline
\multicolumn{1}{|l|}{CPU Type}      & Out of Order & \multicolumn{1}{l|}{L1 Size} & 64 KB (32 KB + 32 KB) & \multicolumn{1}{l|}{Rank Size}         & 2 GB      \\
\multicolumn{1}{|l|}{Number of CPU} & 4            & \multicolumn{1}{l|}{L2 Size} & 512 KB                & \multicolumn{1}{l|}{Row Buffer Policy} & Open Page \\
\multicolumn{1}{|l|}{CPU Frequency} &
  2 GHz &
  \multicolumn{1}{l|}{L3 Size} &
  4 MB &
  \multicolumn{1}{l|}{Scheduling Policy} &
  Rank then Bank Round Robin \\ \hline \hline
\end{tabular}%
}
\end{table}
    For benchmark evaluation, we've used gem5~\cite{gem5} simulator along with DRAMSim2~\cite{dramsim}. For synthesizing, we have used Synopsys Design Compiler with 28 nm technology node. Synthesized components include utilization counters, power gating circuits via sleep transistors, and a power management unit with a re-mapper.
    
    Experiments were carried out in a system specified in Table-\ref{tab:params} with multi-channel memories with sizes of 2 GB and 4 GB. Each DRAM rank contains 8 banks. Irrespective of the number of channels used, we restrict migrations and remapping within the memory channel in order to avoid excessive bandwidth congestion. We aim at measuring energy savings by closing a few affected banks. In each of these experiments, we executed 1 billion instructions across 15 benchmarks selected from the SPEC CPU2006 benchmark suite. 
   
    \subsection{Experiments}
        \label{sec:exp}
        
        The primary motivation of this work is to avoid the usage of slower areas of a DRAM device. All the benefits of VAR-DRAM are a byproduct of the same, including energy savings, reduced latency, higher reliability, and, minimizing the blocking time of the DRAM device. A simulated DDR4 DRAM device would map its results to an ideal DRAM device. Hence, we consider this particular setup as an \textit{Ideal} device henceforth. On the other hand, a PV-affected DRAM device would exhibit incremented key timing parameters in certain areas which would exhibit an increased latency and subsequently higher energy consumption. We term this as \textit{PV-Affected}. Note that a typical DDR4 DRAM device used in a day-to-day scenario resembles with PV-Affected DRAM model~\cite{vampire}.
        
        We compare the energy consumption of VAR-DRAM with PV-Affected and Ideal models. This is done across several configurations. Our PV-aware model incurs additional clock cycles if there are address collisions due to remapping during address translation. In order to study collisions, we design a few composite benchmarks with the aforementioned SPEC CPU2006 benchmarks with synthetic memory traces. The other instance where there will be performance loss would be during data remapping in both dynamic opening and closing of banks, where we would incur an additional latency and bandwidth. We operate our model in two scenarios: (a) the model has foreknowledge of PV affected components as provided by the DRAM manufacturer (static closing), and, (b) PV maps are initialized after the system boots giving up a few clock cycles until PV initialization is reached (dynamic closing). In addition, we also measure the average size of both the trie used. Toward our original motivation of reducing the latency of the DRAM device due to PV, we evaluate the latency of the system in the case of an Ideal, a PV-Worst, and the proposed model. The final analysis would include the area requirement of our remapping logic and its subsequent power requirements. We also perform an analysis on the power gating module installed on each of the banks of the modified DRAM device.

%% file: Text/analysis.tex
\section{Results and Analysis}
    \label{sec:result}
    This section is divided into four sub-sections. The first three largely analyze the results obtained in terms of energy savings, reduction in latency, and refresh count. The final sub-section discusses the overheads associated with VAR-DRAM.
    \subsection{Energy Savings}
        \label{subsec:r_energy}
\begin{figure*}[h]
\begin{center}
\includegraphics[width=0.98\textwidth]{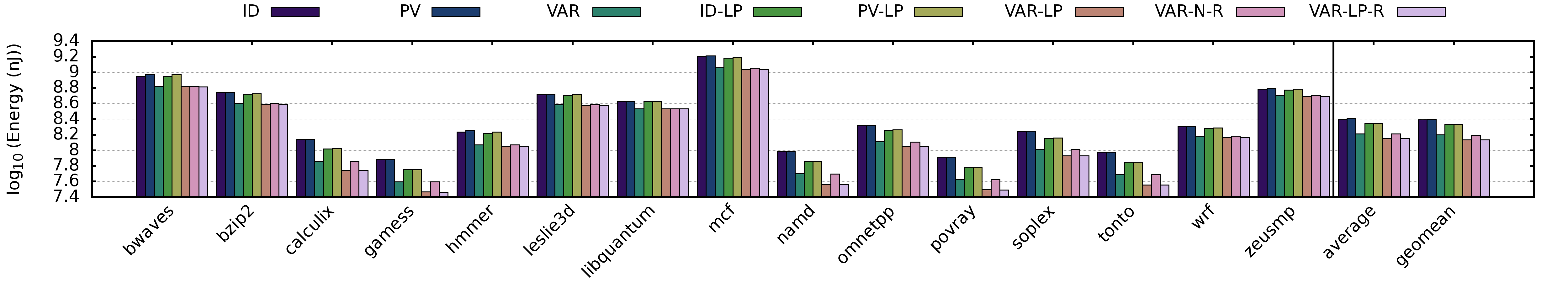}
\caption{Energy Consumption of the Memory Sub-System for 2 GB}
\label{fig:energy-2} 
\end{center}
\end{figure*}

\begin{figure*}[h]
\begin{center}
\includegraphics[width=0.98\textwidth]{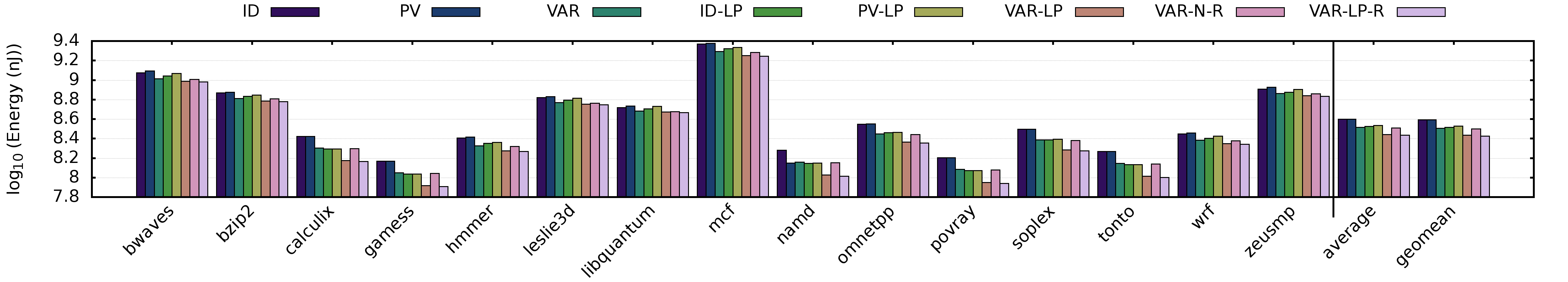}
\caption{Energy Consumption of the Memory Sub-System for 4 GB when 4 Banks were Powered Down}
\label{fig:energy-4.1} 
\end{center}
\end{figure*}

\begin{figure*}[h]
\begin{center}
\includegraphics[width=0.98\textwidth]{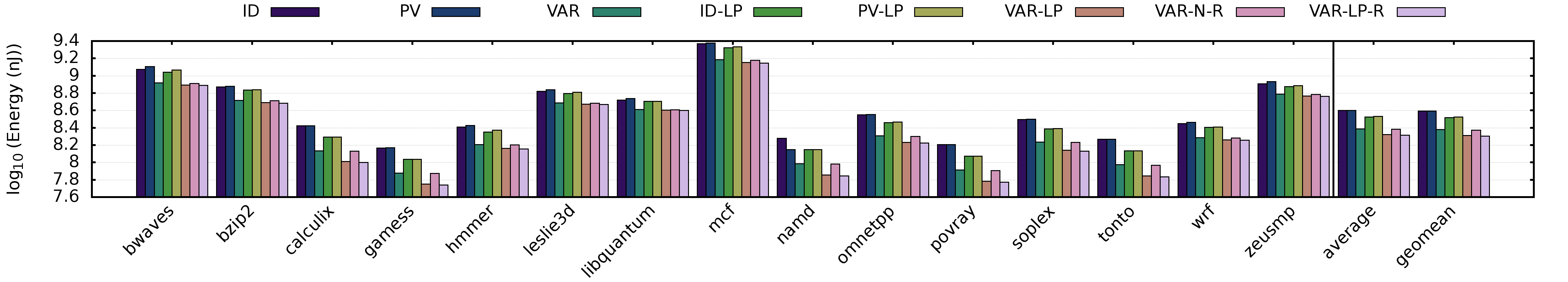}
\caption{Energy Consumption of the Memory Sub-System for 4 GB when 8 Banks were Powered Down}
\label{fig:energy-4.2} 
\end{center}
\end{figure*}

        Energy savings is one of the primary contributions of VAR-DRAM. In this section, we experimentally maintain our claim and show energy savings across different energy-consuming profiles of a DRAM device. We start off with the three basic cases: Ideal (ID), PV-Affected (PV), and, VAR-DRAM (VAR), where we present the total energy consumption of the DRAM device which includes background energy, ACT/PRE energy, burst energy and refresh energy. Alongside, we present the energy profile of all three cases using the most popular DRAM energy-saving technique, i.e., transitioning of ranks into \textit{low power mode (LP)}~\cite{micro,ddr4,dramsim} in respective cases represented by ID-LP, PV-LP, VAR-LP. Finally, we also show energy consumption of VAR-DRAM with reduced refreshing (0.25x REF, tREFW = 256 ms) in both normal (VAR-N-R) and low power modes (VAR-LP-R). Further analysis on VAR-DRAM with reduced refresh instructions is done in Section~\ref{subsec:ref_in_res}. These results are presented in two configurations of memory sizes: 2 GB, depicted in Figure~\ref{fig:energy-2}, and, 4 GB, depicted in Figure~\ref{fig:energy-4.1} and Figure~\ref{fig:energy-4.2} for 4 banks and 8 banks powering down respectively. 

        VAR-DRAM allows several other energy-saving mechanisms to be stacked on top of it, which is evident in LP and R modes. This allows a further saving in energy. Compared to the basic three cases, VAR-DRAM provides energy savings up to 48.81\% in the case of namd in 2 GB, 4 banks powering down case. The amount of savings averages at 29.54\% with a geometric mean of 36.40. Even compared to PV-LP, VAR-DRAM provides an average power savings of 25.78\%. VAR-DRAM when used with LP mode, further provides average energy savings of about 42\%. VAR-DRAM with reduced refreshing (0.25x REF, tREFW = 256 ms) provides average energy savings of 35.59\%. Finally, VAR-DRAM, when stacked up with LP mode and reduced refreshing, provides a maximum average energy savings of 42.13\%. The energy savings is self-explanatory as we are completely powering down DRAM banks, which cuts down both static and dynamic power components. Background energy and refresh energy of powered down banks tend to zero, however, ACT/PRE and burst energy remains the same as all memory accesses for powered down banks are translated and served from their target location. There is a minimal amount of energy added, which is incurred during address translation using the trie.
        
    \subsubsection{Comparative Analysis}
        In this section, we show a comparison of the amount of energy savings obtained in a 2 GB DRAM device for VAR-DRAM (2 banks and 4 banks power-down case) with PV-Affected DRAM. In addition, we also chose to compare our results with a few recent and also other established DRAM energy enhancement techniques, including CROW~\cite{crow}, Flikker~\cite{flikker}, Raha \textit{et al.}~\cite{approx1}, Raha \textit{et al.}~\cite{approx2}, and, RAIDR~\cite{raidr}. 
        CROW uses the concept of copying DRAM rows, similar to VAR-DRAM. The idea of CROW is to make a copy of a given DRAM row in its \textit{CROW-cache}, which is an In-DRAM cache substrate. Simultaneously accessing a row from the primary DRAM array and the CROW-cache reduces access latency of the DRAM device. This process is also known as cell-coupling. Programs are completed faster than their conventional counterparts, thus proving benefits in terms of energy savings. Flikker is another established work, with whom we compare our results. It allows the partitioning of data into either critical or non-critical data. For error-tolerant applications, Flikker then throttles the refresh rate of the non-critical data segment to exploit refresh energy savings. Works by Raha \textit{et al.}~\cite{approx1} and Raha \textit{et al.}~\cite{approx2} are similar to Flikker. The former work, \textit{i.e.},~\cite{approx1} is an extension of the latter work~\cite{approx2} by carefully mitigating \textit{variable retention time} errors. The variable retention time phenomenon was previously discussed in Section~\ref{sec:vrt}. Data in these works are placed into \textit{quality bins} (qbins). The initial qbin, \textit{i.e.,} \textit{qbin0} is free from any errors. Subsequent qbins induces errors on the data, placed on such bins by varying the refresh rate. The aim is to make the DRAM device use varied refresh rates for error tolerant applications, and, as a consequence, obtain energy savings in terms of refresh energy. In the case of Flikker and approximate DRAM refreshing techniques, non-critical data or error tolerant data is usually placed in variation affected regions. RAIDR~\cite{raidr} exploits the fact that not all DRAM rows require refreshing at tREFW. tREFW is set by a DRAM manufacturer in order to account the worst case retention time of variation affected rows. The authors maintain data retention time of all DRAM rows approximately by using bloom filters. 
        In addition to the aforementioned techniques, we have also demonstrated energy savings obtained in a typical low power mode of a modern-day DRAM device. 
        
\begin{figure*}[h]
\begin{center}
\includegraphics[width=0.98\textwidth]{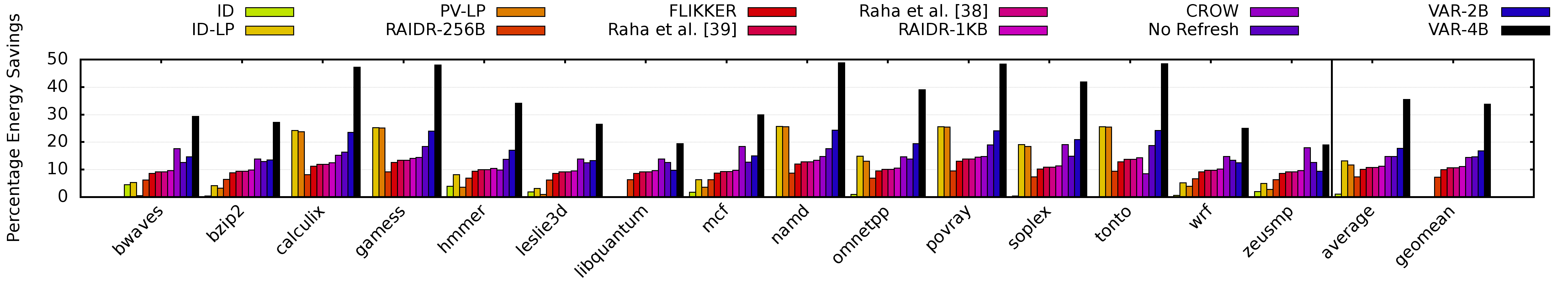}
\caption{Total energy Savings Across Different Energy Saving Mechanisms}
\label{fig:energy_comp} 
\end{center}
\end{figure*}

        We have used DRAMSim2 to implement all of the aforementioned techniques. In the case of Flikker, we have used a lower refresh rate for 3/4\textsuperscript{th} of a DRAM bank. As per Flikker, we emulate non-critical data placement on areas with a lower refresh rate. While implementing all approximate DRAM techniques~\cite{flikker,approx1,approx2}, we assumed that the programs will not be affected by erroneous values, and the execution will complete. In Figure~\ref{fig:energy_comp}, we show the comparison of percentage total energy saved in its respective energy-saving technique. The base case is taken to be a process variation affected DRAM device. Percentage gains in terms of energy are plotted in the figure. The column ID represents energy savings obtained by using an ideal DRAM device with no PV. On average, an ideal DRAM device provides energy benefits of 1.11\%. In addition, low powered mode of an ideal DRAM device (ID-LP) provides energy gains of 13.16\% on average. A PV-affected DRAM device, when operated in low power mode (PV-LP), provides an average energy savings of 11.71\%. CROW, on the other hand, provides an average energy savings of 14.8\% on average across all benchmarks. Flikker shows an energy savings of 10.18\% (FLIKKER). This is due to the aforementioned implementation methodology where critical data only consume a quarter of a DRAM bank. Further, Raha et al.~\cite{approx2} demonstrate energy savings of 10.83\% on average. Its extension (Raha et al.~\cite{approx1}) exploits VRT rows as a means to save energy. Their overall energy savings was 10.8\%. RAIDR was evaluated in two different bloom filter sizes: 256 B (RAIDR-256B) and 1 KB (RAIDR-1KB). RAIDR-256B saved 7.42\% of the total DRAM device's energy whereas RAIDR-1KB had a significantly higher energy savings value, \textit{i.e.}, 10.18\% on average. While RAIDR mostly can accommodate all types of applications: error-tolerant or otherwise, the approximation methods cannot. VRT rows cannot be profiled using a static retention-time profiling technique like RAIDR. When 2 banks out of 8 banks (2 GB module) are powered down in the case of VAR-DRAM, we achieve an average energy savings of 17.1\% (VAR-2B). Finally, we see that VAR-DRAM, when operated at 4 banks out of 8 (VAR-4B), achieves the highest energy savings across all the techniques for each benchmark, which further achieves a higher percentage of energy savings which averages at 35.51\%. 

        Most of these aforementioned works are aimed at reducing refresh energy only. In case of extended refresh window times, the device is configured with a tREFW time equivalent to 256 ms (4x), instead of the standard 64 ms. RAIDR-256B saves 49\% refresh energy on average. Further, the other variant of RAIDR, \textit{i.e.} RAIDR-1KB saves 73.65\% of refresh energy on average. The value is slightly lower than the theoretically expected value of 50\% or 75\% respectively due to the inclusion of false positives, introduced via bloom filters. Three-fourth of a Flikker bank was configured to use 256 ms (4x) during refreshing, while the remaining portion of the DRAM bank was refreshed at the standard 64 ms tREFW time. We observed refresh energy savings of around 68.6\%. The methods proposed by Raha \textit{et al.}~\cite{approx1} and Raha \textit{et al.}~\cite{approx2} induces errors on the DRAM device. Although these techniques are accuracy-configurable, however, it is unlikely that these methods can save energy in case of applications having a significant percentage of critical data. Overall, we observed average refresh energy savings of 73\% and 72.7\% respectively. CROW shows similar refresh energy savings as VAR-DRAM. This is due to the fact the CROW uses its CROW-cache to remap weak rows and refreshes the remaining array at 256 ms tREFW. VAR-DRAM, on the other hand, remaps these weak rows to strong rows and refreshes the entire DRAM array at 256 ms. Since CROW uses the CROW-cache, therefore, the overall static power requirements of the same increase. This directly increases the background energy as the effective array size is increased, thus requiring a higher operating current. Introducing errors largely limit the application of such hardware devices, which is seen in cases of~\cite{flikker,approx1,approx2}. 
        
        In most of these aforementioned energy-saving techniques, we ignore the largest energy consumer of the DRAM device, \textit{i.e.}, background energy. Background energy consumes around 65\% of the total DRAM device's energy~\cite{dramsim}. Static energy is one of the dominant battery-draining factors in the case of mobile devices. VAR-DRAM addresses this issue. In addition, it also addresses the excessive DRAM refreshing issue.  

    \subsection{Timing Analysis}
        We have sub-divided this section into two parts for a deeper analysis: (a) access latency, and, (b) total execution time of a program. This helps us to differentiate between the speedup gains obtained in the memory sub-system and the entire system.  
        \subsubsection{Access Latency}
            \label{sec:r_acclat}
\begin{figure*}[h]
\begin{center}
\includegraphics[width=0.98\textwidth]{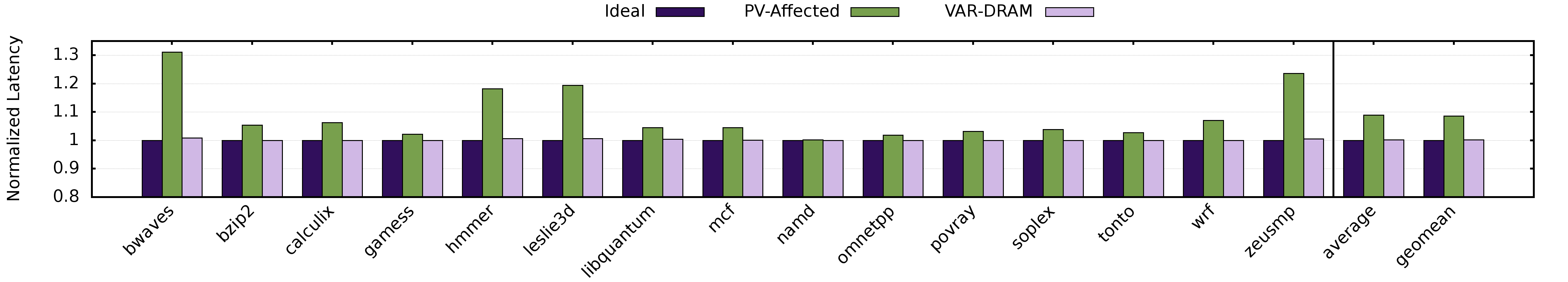}
\caption{Normalized Access Latency of the Memory Sub-System}
\label{fig:acc_lat} 
\end{center}
\end{figure*}

            Section~\ref{sec:acclat} already defines access latency. In our experiment, we measure the access latency of each DRAM bank while accounting for additional time incurred during address translation and data remapping. VAR-DRAM operates at a lower capacity of the DRAM banks alongside its usage of faster DRAM areas. Address translation during collisions and data remapping adds a significant amount of overhead in terms of latency. In Figure~\ref{fig:acc_lat} we demonstrate the aforementioned experiment with a comparison between Ideal, PV-Affected, and VAR-DRAM. Even stacked with additional overheads, VAR-DRAM is likely to exhibit lower latency terms than PV-Affected, which it does with 7.4\% lower access latency on average. VAR-DRAM only deviates from a theoretical DRAM device (Ideal) by a marginal 0.8\% on average. We did a worst-case analysis of access latency for VAR-DRAM. The memory controller is patched to send an \textit{INTERRUPT} signal for every translation it makes. This effectively increases tRAS time of memory access falling into a victim bank. We have uniformly increased tRAS time for all victim addresses for this analysis.
            
        \subsubsection{Execution Time}
            \label{sec:r_exec}
\begin{figure*}[h]
\begin{center}
\includegraphics[width=0.98\textwidth]{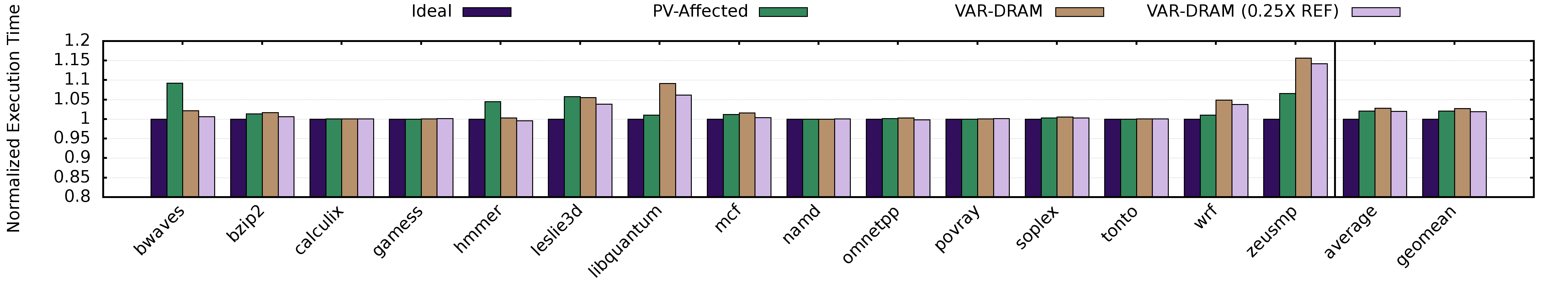}
\caption{Normalized Execution Time of the System}
\label{fig:exec_time} 
\end{center}
\end{figure*}

            In Section~\ref{subsec:m_latency}, we already explained that a reduction in memory latency does not always imply a reduction in total execution time. In our case, since the number of row decoders are reduced, the overall execution time may be affected, depending upon the memory access pattern of a given benchmark. Memory intensive benchmarks like bwaves show an overall reduction in execution time by 7.5\% over the PV-Affected case. Whereas, benchmarks like libquantum, mcf and zeusmp show an overall increment in the total execution time. In Figure~\ref{fig:exec_time}, we report the normalized execution time of all the benchmarks over its Ideal case. There are 3 other cases shown in that experiment, including PV-Affected, VAR-DRAM, and VAR-DRAM with refresh instructions reduced by 75\% while remapping all the weak rows to a healthy location and accounting for its latency. VAR-DRAM shows an overall increment over the PV-Affected case by 0.7\% additional time. VAR-DRAM, when used alongside reduced refreshes, provides an overall reduction in execution time by approximately 0.05\%. Note that this is the worst-case execution time of reduced refresh, as we remap all weak rows at the start. 
      
    \subsection{Reduction in Number of Refreshes}
        \label{subsec:ref_in_res}
\begin{figure*}[h]
            \begin{center}
            \mbox{
            \hspace{-1.0ex}
            \subfigure[Number of Refresh Instructions Issued by the Memory Controller]
                {
                \label{fig:ref_count}
                \includegraphics[width=0.48\textwidth]{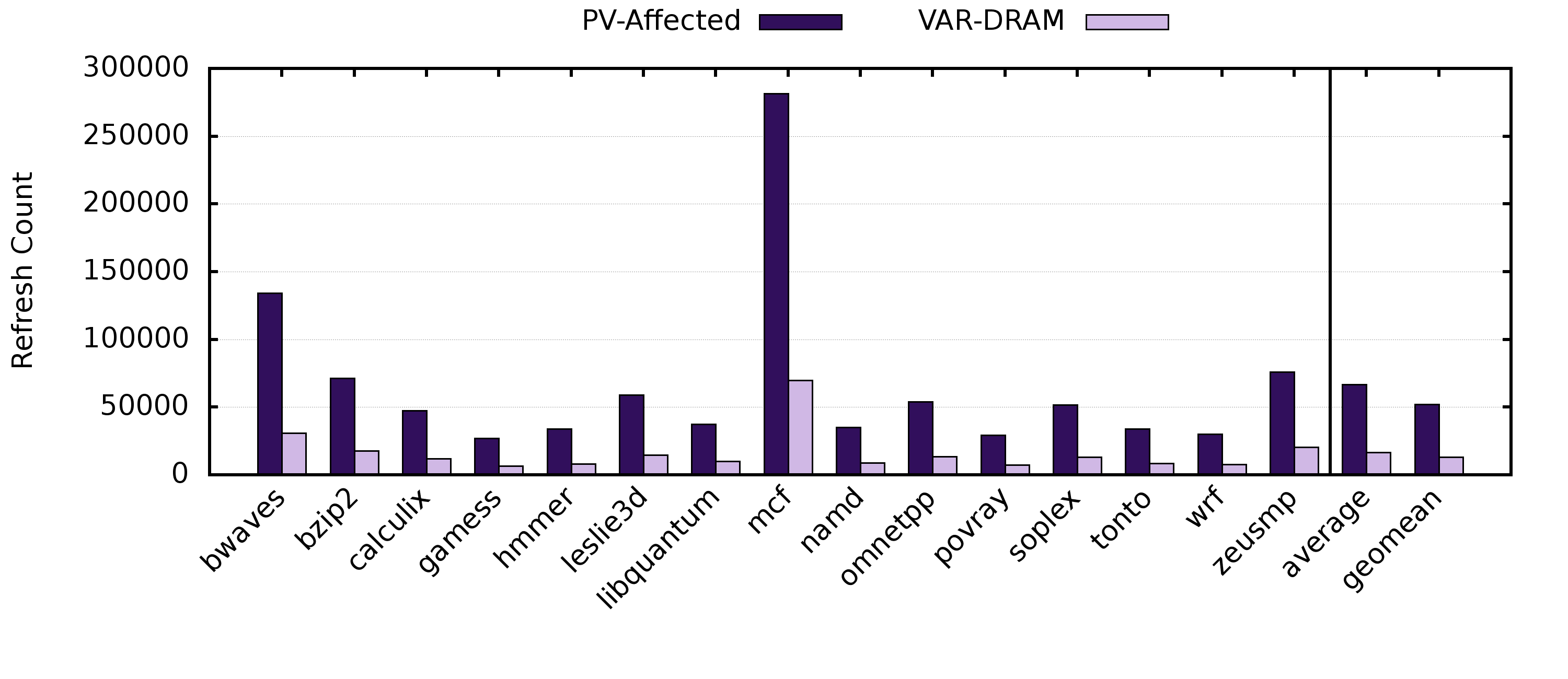}
                }
            \hspace{-1.0ex}
            \subfigure[Refresh Energy consumed by the Memory Sub-Section]
                {
                \label{fig:ref_energy}
                \includegraphics[width=0.48\textwidth]{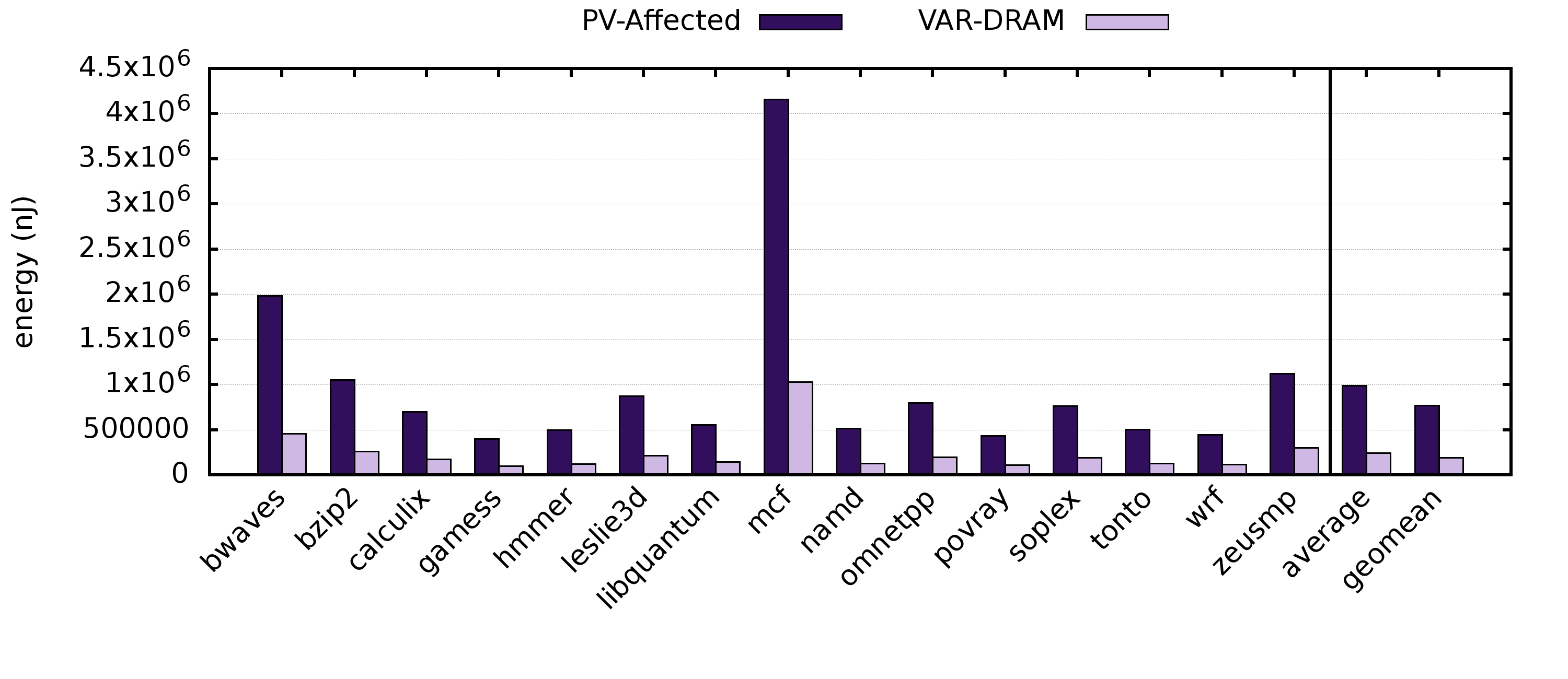}
                
                }
            }
            \caption{Analysis on Refresh Instruction Reducing Capabilities of VAR-DRAM}
            \label{fig:ref_all} 
            \end{center}
        \end{figure*}

        Section~\ref{subsec:m_retention} already states the fact that VAR-DRAM can be stacked over existing DRAM technologies to remap weak rows and reduce refresh instruction count. This minimizes the blocking time of the device, which is previously stated in Section~\ref{sec:r_exec} and shown in Figure~\ref{fig:exec_time}. Figure~\ref{fig:ref_count} further shows the reduction in the number of refresh instructions reduced in our experiments (PV-Affected case compared with VAR-DRAM). Alongside, the amount of additional time incurred for remapping for each benchmark is represented in Figure~\ref{fig:ref_overhead}. Figure~\ref{fig:ref_energy} shows the amount of refresh energy consumed in both PV-Affected and VAR-DRAM cases. In either of the cases, the amount of reduction is 75.2\%. The average overhead incurred during remapping is 8.7 $\mu$s. In Section~\ref{sec:r_exec}, we have already shown that inclusive of this overhead, on average, we achieve faster execution time over PV-Affected case.

    \subsection{Overhead}
        \label{subsec:overhead}
        We categorize overheads associated with VAR-DRAM into four categories: (a) performance, (b) storage, (c) area, and (d) additional power. We discuss each of these overheads in this sub-section.

        \subsubsection{Performance}
\begin{figure*}[h]
            \begin{center}
            \mbox{
            \hspace{-1.0ex}
            \subfigure[Extra Time Required due to Remapping of Weak Rows]
                {
                \label{fig:ref_overhead}
                \includegraphics[width=0.48\textwidth]{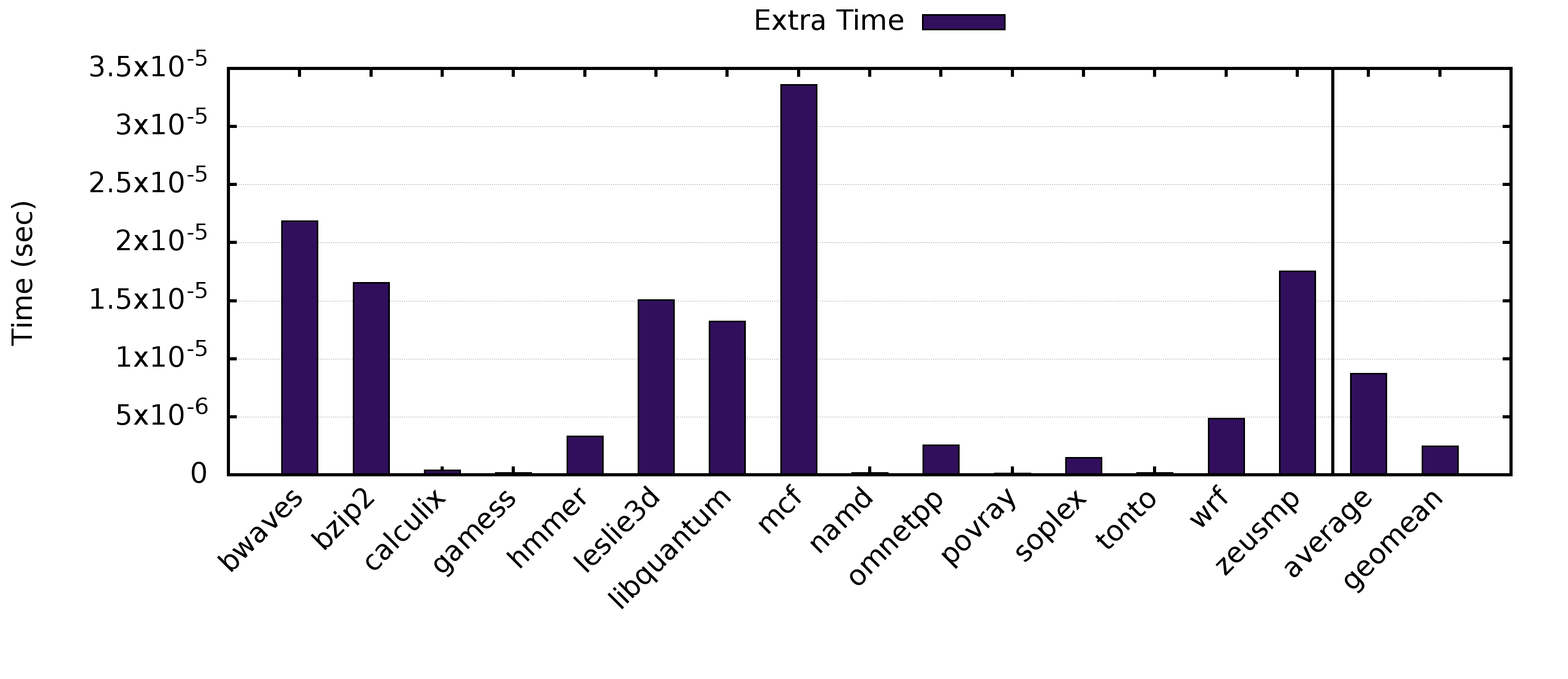}
                }
            }
            \hspace{-1.0ex}
            \subfigure[Translation Overhead on Synthetic Benchmarks]
                {
                \label{fig:t_ov}
                \includegraphics[width=0.48\textwidth]{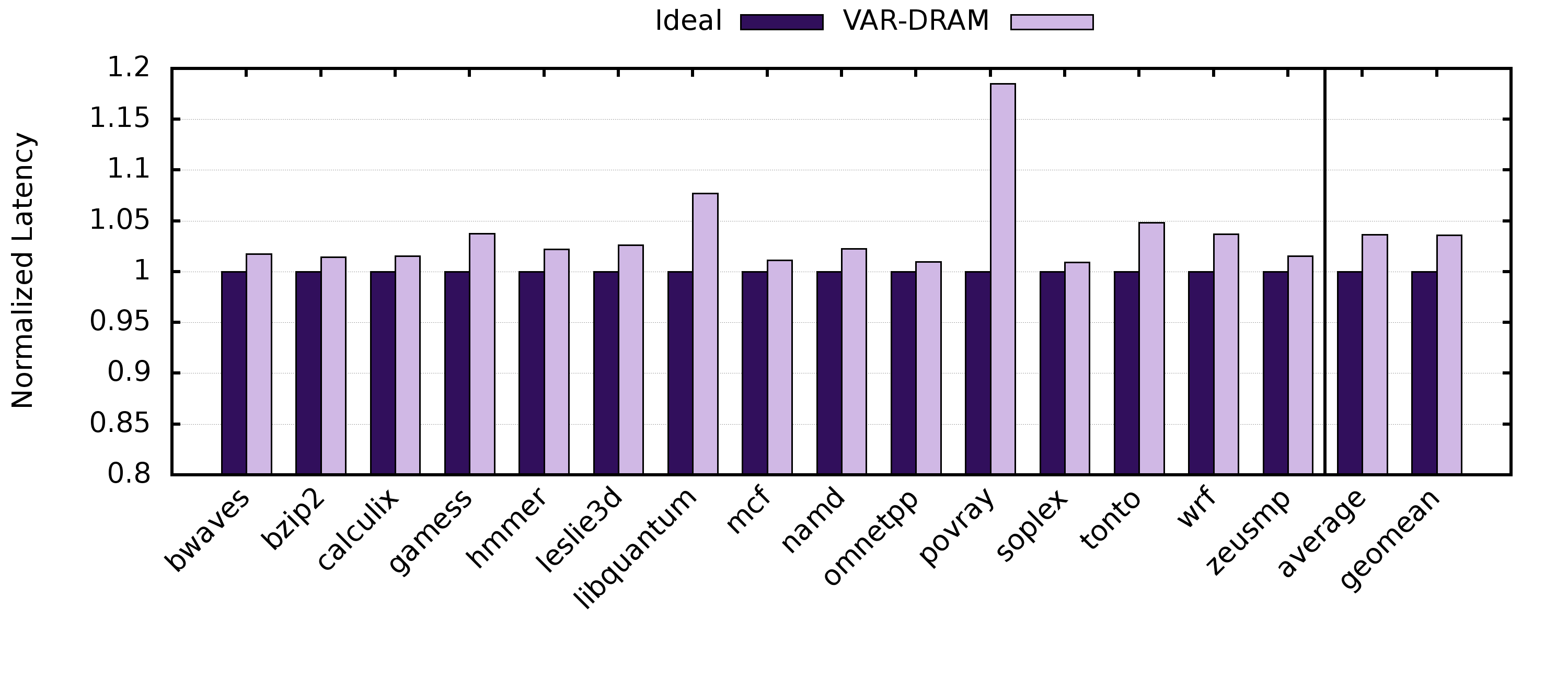}
                }
            \caption{Analysis on Weak Rows and Address Collision in VAR-DRAM}
            \label{fig:ref_all_x} 
            \end{center}
        \end{figure*}

        In Section~\ref{sec:method}, we already discussed the instances where VAR-DRAM is likely to incur performance overhead. During address collision, there will be additional 3 stall cycles, required to traverse the trie to find the correct address. During evaluation using SPEC CPU2006 benchmarks, we did not encounter address collision while allocating target addresses corresponding to victim addresses. Therefore, we created separate synthetic memory traces for each benchmark, and using these composite benchmarks, we show translation overhead. Figure~\ref{fig:t_ov} shows the amount of extra time required for translating addresses, which averages at 3.6\% compared to an ideal DRAM device. 
        \subsubsection{Bandwidth, Migration and Remapping}
        As for latency incurred during data remapping, Figure~\ref{fig:acc_lat} in Section~\ref{sec:r_acclat} already demonstrates the same. During data migration, we are also likely to incur additional bandwidth. Bandwidth is a precious commodity in DRAM devices~\cite{rowclone,intelpaper}. By using RowClone and restricting inter-channel data remapping, we limit this overhead. In the benchmarks that we have used for experimentation, the entire memory bandwidth is never fully utilized. In fact, only programs with memory requests after every tRC time, ideally a cache-less system, would utilize or over-utilize the full extent of available bandwidth. This factor helps us to establish the fact that migration and remapping of addresses in the proposed experiments have a minute effect on the overall bandwidth of the memory system.

\begin{figure*}[t]
\begin{center}
\includegraphics[width=0.98\textwidth]{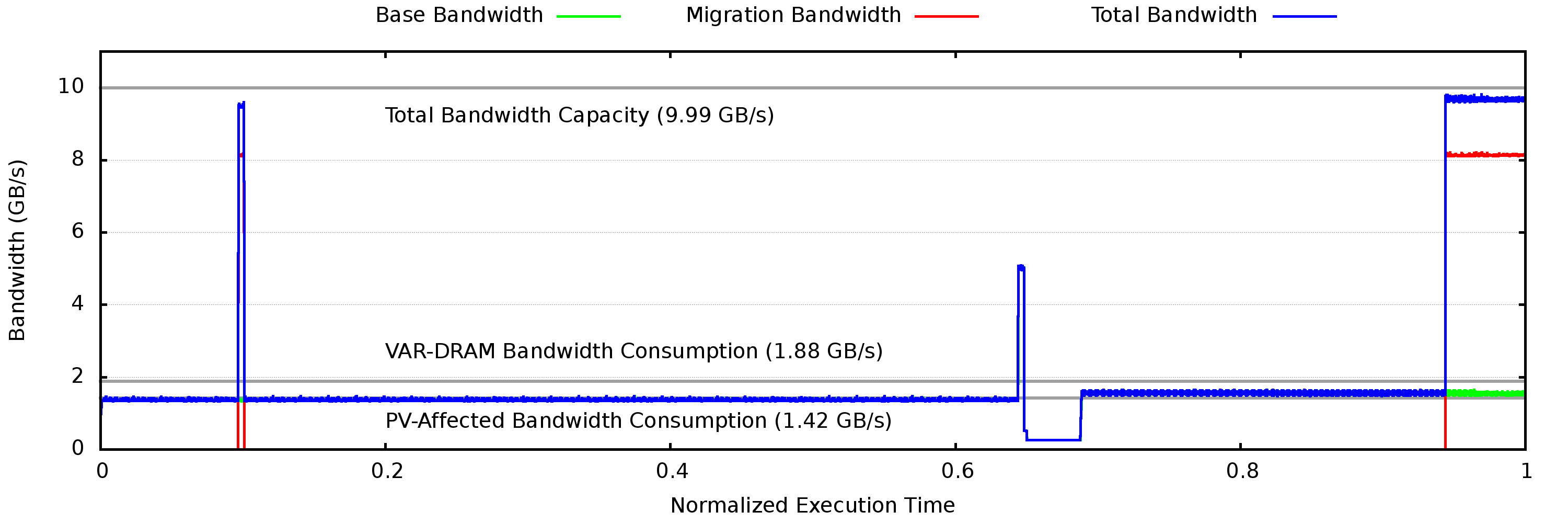}
\caption{Bandwidth Case Study for \textit{bwaves}}
\label{fig:case_bwaves} 
\end{center}
\end{figure*} 
        
\begin{figure*}[h]
            \begin{center}
            \mbox{

            \hspace{-1.0ex}
            \subfigure[Normalized Bandwidth Usage]
                {
                \label{fig:b_ov}
                \includegraphics[width=0.48\textwidth]{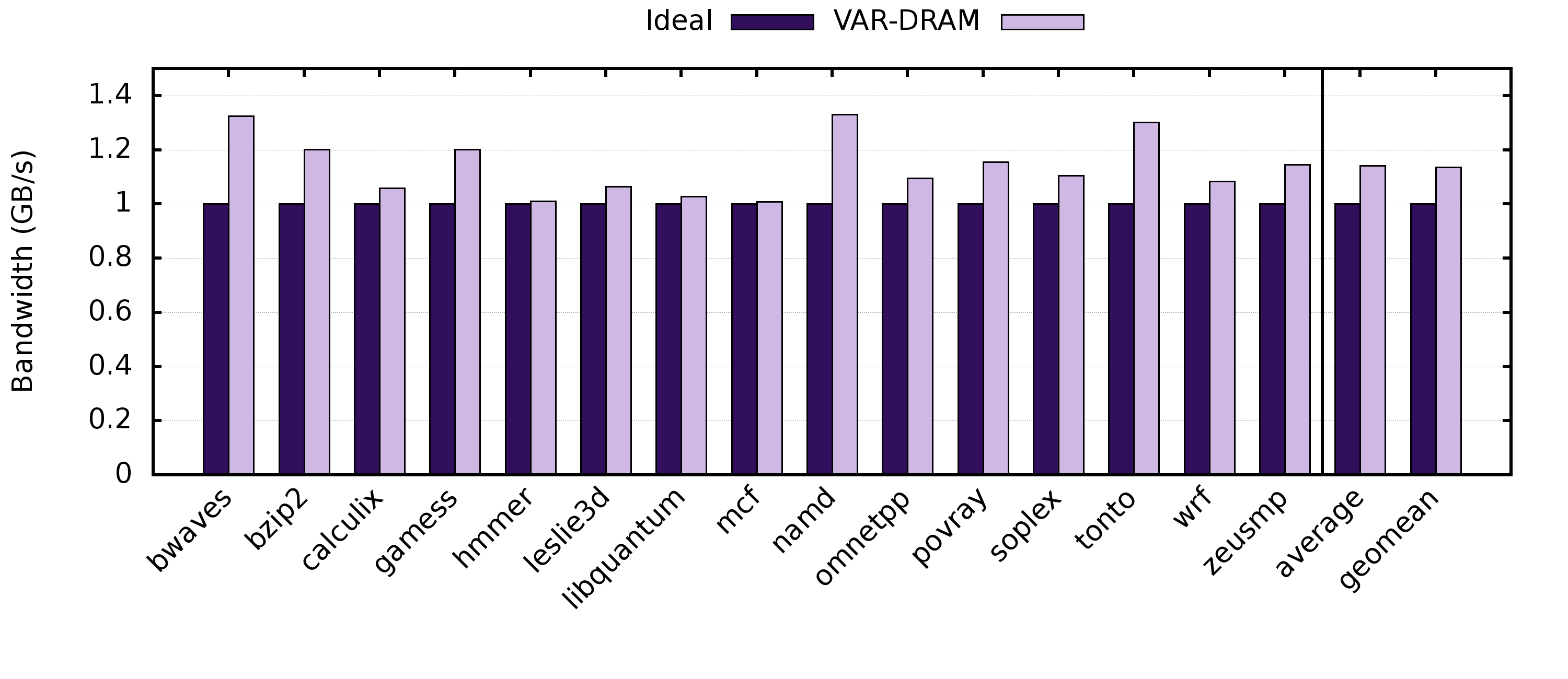}
                
                }
            \hspace{-1.0ex}
            \subfigure[Trie Size (Both Primary Trie as well as the Auxiliary Trie)]
                {
                \label{fig:t_size}
                \includegraphics[width=0.48\textwidth]{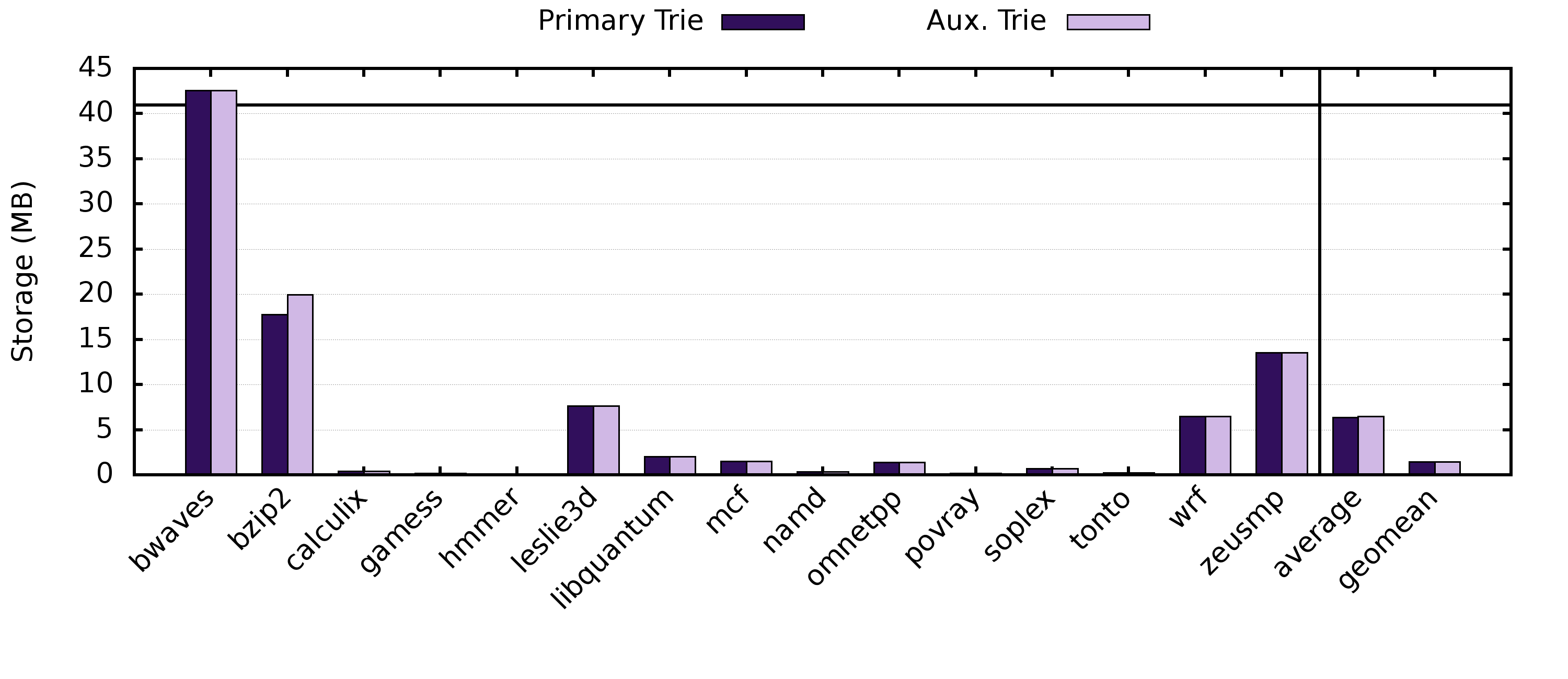}
                }
            }
            \caption{Analysis on Bandwidth and Storage Overhead incurred during Evaluation of VAR-DRAM}
            \label{fig:ov_all} 
            \end{center}
        \end{figure*}

        In Figure~\ref{fig:case_bwaves}, we demonstrate the bandwidth consumption of a particular benchmark \textit{bwaves} as a special case. The bandwidth consumption is averaged at 100,000 clock cycles. From Figure~\ref{fig:t_size}, it is clear that the program initiates a back migrate signal as a precaution of the memory size running out. For this particular experiment, we have first forward migrated a set of PV affected addresses from 4 DRAM banks to their respective healthy bank locations at 9.65\% of the program's execution time. During remapping of all PV affected rows, there is an instantaneous bandwidth consumption, which can be regulated by keeping a time $>$ tRC between two remapping addresses. In this case, we have considered that all addresses are issued migration requests at the same time. All banks reopen at 94.3\% of the program, initiating another remapping call. Migration bandwidth is highlighted as a red line, while the base bandwidth is highlighted by a green line. Overall bandwidth consumption is highlighted by a blue line. The total available bandwidth of the system is 9.99 GB/s, while VAR-DRAM consumes 1.88 GB/s, which is only a fraction of the total available bandwidth. The normal execution of \textit{bwaves} consumes 1.42 GB/s bandwidth. Overall, VAR-DRAM consumes 32.3\% more bandwidth than its normal counterpart. \textit{bwaves} has the biggest pool of unique addresses, which is further explained in the following section. Therefore, in terms of bandwidth, the memory system is never taxed to the full capacity of its bandwidth. Figure~\ref{fig:b_ov} further highlights the bandwidth requirements of each of the benchmarks, which averages at 14.07\% overhead, when data migration and remapping is forced during the final tenth of the program's execution time.

        As previously mentioned in Section~\ref{sec:method}, migration and remapping are performed in the background. Normal memory requests are served during this phase. This can also be seen in Figure~\ref{fig:case_bwaves}. There are two migration and remapping calls during the runtime of the program. The overall bandwidth consumption rate (blue line) reflects the normal, and as well as data remapping bandwidth, when performed simultaneously. Instances of overlapped read or write requests appear, however, such requests are properly handled as mentioned in Section~\ref{sec:method}.

        
        
        \subsubsection{Storage}
            \label{sto_ov_x}
            Although the growth of the trie is dynamic, implementation in Silicon would require us to allocate a fixed size for the trie. Note that we have two trie structures: (a) one acting as a translation table (called \textit{primary trie}), and, (b) the other is a companion trie, called \textit{auxiliary trie}, used for preserving the state of the DRAM device during data remapping. During all our experiments, we have used an upper limit of both the trie at 2\% of the total memory size. Since a trie is more space-efficient than a normal table, therefore we are able to store more than 2 million addresses in the companion trie, and, half of that in the translation trie in the case of a 2 GB DRAM device. All powered down banks open up when 90\% of either of the trie space is consumed. This is periodically checked in parallel to the normal DRAM operations. A signal for bank re-opening is sent to the memory controller. The control unit then proceeds to turn on banks. Additionally, we refrain from powering down any more banks until the program completes.
            
        \subsubsection{Area}
            We previously discussed additional hardware components, required for VAR-DRAM, in Section~\ref{sec:hw_impl_para}. Our synthesized results obtained using Synopsys Design Compiler gave a value of 961.01 $\mu$m$^2$ of area including the re-mapper unit, utilization computation unit, and sleep transistors for banks in one rank. The sleep transistors exclusively consume 19.5 $\mu$m$^2$ of area. Compared to a 400 mm$^2$ chip, our additional hardware components require 0.002\% area of the chip.  
        \subsubsection{Additional Power Overhead}
            \label{sec:power-ov}
            Our model when synthesized via Synopsys Design Compiler, reports that the sleep transistors consume 8.89 nW of static power. One controller with a utilization computation unit consumes about 42.84 $\mu$W of power. The power gated implementation has a transient energy consumption of about 1.2 pJ. This is spent when the circuit changes its state from sleep to wake up mode.      

%% file: Text/conclusion.tex
\section{Conclusion and Future Work}
    \label{sec:conclusion}
    In this work, we have proposed a variation aware DRAM framework that efficiently re-configures the energy utilization of variation affected DRAM devices. Along with saving energy, the framework reduces memory access latency and execution time of programs on average via the remapping of addresses. Our initial claims of saving energy, reducing memory latency, and preserving the state of the DRAM device is held in Section~\ref{sec:result}, where we have shown 29.54\% energy savings on average over a PV-Affected DRAM device while also reducing the access latency of the memory device by 7.4\% on average. An OS-transparent paging policy, if integrated with VAR-DRAM along with compiler support, the amount of energy savings is likely to increase. We plan on working on this in the near future. 

    The reason we call VAR-DRAM a \textit{framework} is that VAR-DRAM is agnostic to the DRAM technology. It can be incorporated with memory technologies, including but not limited to DDR3, DDR4, LPDDRx etc. LPDDRx memories, popular with low-powered devices, already have a feature for a per-bank refresh. A per-bank power-down mode would additionally help to reduce the amount of energy consumed by these devices, thus extending the longevity of the battery. As previously mentioned in Section~\ref{sec:related}, emerging memories and non-volatile memory technologies are also affected by variations. A similar phenomenon is also shown in SSD devices. Thererfore, an extension of VAR-DRAM can provide a variation-aware platform to obtain these device's theoretical parameters. The \textit{lightweight} address remapping logic, partly due to efficient space requirements and logarithmic lookup times, can be used in several other applications. 
    With the shrinking of technologies, variations will be further prevalent in future memories. Therefore, we believe, VAR-DRAM will be beneficial for the design of future memory technologies and encourage researchers to adopt this framework.
    
    

%% file: template.bbl
\begin{thebibliography}{10}

\bibitem{ware}
M.~{Ware}, K.~{Rajamani}, M.~{Floyd}, B.~{Brock}, J.~C. {Rubio}, F.~{Rawson},
  and J.~B. {Carter}.
\newblock {Architecting for power management: The IBM® POWER7™ approach}.
\newblock In {\em HPCA - 16 2010 The Sixteenth International Symposium on
  High-Performance Computer Architecture}, Jan 2010.

\bibitem{hoelze}
I.~{Paul}, W.~{Huang}, M.~{Arora}, and S.~{Yalamanchili}.
\newblock Harmonia: Balancing compute and memory power in high-performance
  gpus.
\newblock In {\em 2015 ACM/IEEE 42nd Annual International Symposium on Computer
  Architecture (ISCA)}, pages 54--65, 2015.

\bibitem{barosso}
Luiz~André Barroso, Jimmy Clidaras, and Urs H\"{o}lzle.
\newblock {The Datacenter as a Computer: An Introduction to the Design of
  Warehouse-Scale Machines, Second edition}.
\newblock {\em Synthesis Lectures on Computer Architecture}, 2013.

\bibitem{micro}
Micron Technology.
\newblock {TN-41-01: Calculating Memory System Power for DDR3}.
\newblock Technical Report {TN41\_01DDR3}, January 2007.

\bibitem{ddr4}
{Micron Technology,}.
\newblock {DDR4 SDRAM}.
\newblock Technical Report {MT40A2G4, MT40A1G8, MT40A512M16}, 2015.

\bibitem{microLPDDRx}
Micron Technology.
\newblock {TN-46-12: Mobile DRAM Power-Saving Features and Power Calculations}.
\newblock Technical Report {TN46\_12}, 2005.

\bibitem{sarangi}
S.~R. {Sarangi}, B.~{Greskamp}, R.~{Teodorescu}, J.~{Nakano}, A.~{Tiwari}, and
  J.~{Torrellas}.
\newblock Varius: A model of process variation and resulting timing errors for
  microarchitects.
\newblock {\em IEEE Transactions on Semiconductor Manufacturing}, 21(1):3--13,
  2008.

\bibitem{process}
B.~Zhao, Y.~Du, J.~Yang, and Y.~Zhang.
\newblock {Process Variation-Aware Nonuniform Cache Management in a 3D
  Die-Stacked Multicore Processor}.
\newblock {\em IEEE Transactions on Computers}, 62(11):2252--2265, Nov 2013.

\bibitem{delaluz}
Victor Delaluz, Mahmut Kandemir, N.~Vijaykrishnan, Anand Sivasubramaniam, and
  Mary~Jane Irwin.
\newblock {Hardware and Software Techniques for Controlling DRAM Power Modes}.
\newblock {\em IEEE Transactions on Computers}, November 2001.

\bibitem{delaluz-bank}
V.~{Delaluz}, M.~{Kandemir}, N.~{Vijaykrishnan}, A.~{Sivasubramaniam}, and
  M.~J. {Irwin}.
\newblock {DRAM energy management using software and hardware directed power
  mode control}.
\newblock In {\em Proceedings HPCA Seventh International Symposium on
  High-Performance Computer Architecture}, Jan 2001.

\bibitem{ras_timing}
{Sang-Hoon Lee}, {Chang-Hoon Choi}, {Jeong-Taek Kong}, {Won-Seong Lee}, and
  {Jei-Hwan Yoo}.
\newblock An efficient statistical analysis methodology and its application to
  high-density drams.
\newblock In {\em 1997 Proceedings of IEEE International Conference on Computer
  Aided Design (ICCAD)}, pages 678--683, 1997.

\bibitem{raidr}
Jamie Liu, Ben Jaiyen, Richard Veras, and Onur Mutlu.
\newblock {RAIDR: Retention-Aware Intelligent DRAM Refresh}.
\newblock In {\em Proceedings of the 39th Annual International Symposium on
  Computer Architecture}, volume~40, New York, NY, USA, 2012. Association for
  Computing Machinery.

\bibitem{lebeck}
Alvin~R. Lebeck, Xiaobo Fan, Heng Zeng, and Carla Ellis.
\newblock Power aware page allocation.
\newblock In {\em Proceedings of the Ninth International Conference on
  Architectural Support for Programming Languages and Operating Systems},
  ASPLOS IX, page 105–116, New York, NY, USA, 2000. Association for Computing
  Machinery.

\bibitem{delaluz-kande}
V.~De~La Luz, M.~Kandemir, and I.~Kolcu.
\newblock {Automatic data migration for reducing energy consumption in
  multi-bank memory systems}.
\newblock In {\em Proceedings Design Automation Conference (DAC)}, June 2002.

\bibitem{crow}
Hasan Hassan, Minesh Patel, Jeremie~S. Kim, A.~Giray Yaglikci, Nandita
  Vijaykumar, Nika~Mansouri Ghiasi, Saugata Ghose, and Onur Mutlu.
\newblock Crow: A low-cost substrate for improving dram performance, energy
  efficiency, and reliability.
\newblock In {\em Proceedings of the 46th International Symposium on Computer
  Architecture}, ISCA ’19, page 129–142, New York, NY, USA, 2019.
  Association for Computing Machinery.

\bibitem{10.1109}
D.~{Shin} and J.~W. {Lee}.
\newblock Bandwidth-aware dram page migration for heterogeneous mobile memory
  systems.
\newblock In {\em 2018 IEEE International Conference on Consumer Electronics
  (ICCE)}, pages 1--5, 2018.

\bibitem{liu2010}
S.~Liu, Y.~Zhang, S.~Ogrenci Memik, and G.~Memik.
\newblock {An Approach for Adaptive DRAM Temperature and Power Management}.
\newblock {\em IEEE Transactions on Very Large Scale Integration (VLSI)
  Systems}, April 2010.

\bibitem{leecho}
H.~{Lee}, S.~{Cho}, and B.~R. {Childers}.
\newblock Performance of graceful degradation for cache faults.
\newblock In {\em IEEE Computer Society Annual Symposium on VLSI (ISVLSI '07)},
  pages 409--415, 2007.

\bibitem{isqed}
K.~{Goswami}, H.~K. {Mondal}, S.~{Das}, and D.~S. {Banerjee}.
\newblock {State Preserving Dynamic DRAM Bank Re-Configurations for Enhanced
  Power Efficiency}.
\newblock In {\em 20th International Symposium on Quality Electronic Design
  (ISQED)}, March 2019.

\bibitem{bank-sensitive-model}
M.~{Jung}, D.~M. {Mathew}, É.~F. {Zulian}, C.~{Weis}, and N.~{Wehn}.
\newblock {A new bank sensitive DRAMPower model for efficient design space
  exploration}.
\newblock In {\em 26th PATMOS}, Sep. 2016.

\bibitem{drampower}
Chandrasekar Karthik, Weis Christian, Li~Yonghui, Goossens Sven, Jung Matthias,
  Naji Omar, Akesson Benny, Wehn Norbert, and Goossens Kees.
\newblock {DRAMPower: Open-source DRAM Power \& Energy Estimation Tool}, 2014.

\bibitem{dramsim}
P.~{Rosenfeld}, E.~{Cooper-Balis}, and B.~{Jacob}.
\newblock Dramsim2: A cycle accurate memory system simulator.
\newblock {\em IEEE Computer Architecture Letters}, 10(1):16--19, 2011.

\bibitem{o34}
T.~{Hamamoto}, S.~{Sugiura}, and S.~{Sawada}.
\newblock {On the retention time distribution of dynamic random access memory
  (DRAM)}.
\newblock {\em IEEE Transactions on Electron Devices}, June 1998.

\bibitem{nvm}
Jalil Boukhobza, St\'{e}phane Rubini, Renhai Chen, and Zili Shao.
\newblock Emerging nvm: A survey on architectural integration and research
  challenges.
\newblock {\em ACM Trans. Des. Autom. Electron. Syst.}, 23(2), November 2017.

\bibitem{stt-pv}
N.~{Sayed}, S.~M. {Nair}, R.~{Bishnoi}, and M.~B. {Tahoori}.
\newblock Process variation and temperature aware adaptive scrubbing for
  retention failures in stt-mram.
\newblock In {\em 2018 23rd Asia and South Pacific Design Automation Conference
  (ASP-DAC)}, pages 203--208, 2018.

\bibitem{other_paper}
Yixin Luo, Saugata Ghose, Yu~Cai, Erich~F. Haratsch, and Onur Mutlu.
\newblock Improving 3d nand flash memory lifetime by tolerating early retention
  loss and process variation.
\newblock {\em SIGMETRICS Perform. Eval. Rev.}, 2018.

\bibitem{vampire}
Saugata Ghose, Abdullah~Giray Yaglik\c{c}i, Raghav Gupta, Donghyuk Lee, Kais
  Kudrolli, William~X. Liu, Hasan Hassan, Kevin~K. Chang, Niladrish Chatterjee,
  Aditya Agrawal, Mike O'Connor, and Onur Mutlu.
\newblock {What Your DRAM Power Models Are Not Telling You: Lessons from a
  Detailed Experimental Study}.
\newblock {\em Proc. ACM Meas. Anal. Comput. Syst.}, 2(3), 2018.

\bibitem{sample}
S.~{Cha}, O.~{Seongil}, H.~{Shin}, S.~{Hwang}, K.~{Park}, S.~J. {Jang}, J.~S.
  {Choi}, G.~Y. {Jin}, Y.~H. {Son}, H.~{Cho}, J.~H. {Ahn}, and N.~S. {Kim}.
\newblock {Defect Analysis and Cost-Effective Resilience Architecture for
  Future DRAM Devices}.
\newblock In {\em IEEE HPCA}, pages 61--72, 2017.

\bibitem{ozturk}
Ozcan Ozturk and Mahmut Kandemir.
\newblock Ilp-based energy minimization techniques for banked memories.
\newblock {\em ACM Trans. Des. Autom. Electron. Syst.}, 13(3), July 2008.

\bibitem{wang}
Y.~Wang, L.~Zhang, Y.~Han, H.~Li, and X.~Li.
\newblock {Data Remapping for Static NUCA in Degradable Chip Multiprocessors}.
\newblock {\em IEEE Transactions on Very Large Scale Integration (VLSI)
  Systems}, May 2015.

\bibitem{lu2016}
Y.~Lu, D.~Wu, B.~He, X.~Tang, J.~Xu, and M.~Guo.
\newblock {Rank-Aware Dynamic Migrations and Adaptive Demotions for DRAM Power
  Management}.
\newblock {\em IEEE Transactions on Computers}, Jan 2016.

\bibitem{lee2017}
Y.~Lee, H.~Kim, S.~Hong, and S.~Kim.
\newblock {Partial Row Activation for Low-Power DRAM System}.
\newblock In {\em 2017 IEEE International Symposium on High Performance
  Computer Architecture (HPCA)}, Feb 2017.

\bibitem{reduced_volt}
Kevin~K. Chang, A.~Giray Ya\u{g}l\i{}k\c{c}\i{}, Saugata Ghose, Aditya Agrawal,
  Niladrish Chatterjee, Abhijith Kashyap, Donghyuk Lee, Mike O’Connor, Hasan
  Hassan, and Onur Mutlu.
\newblock Understanding reduced-voltage operation in modern dram devices:
  Experimental characterization, analysis, and mechanisms.
\newblock {\em Proc. ACM Meas. Anal. Comput. Syst.}, 2017.

\bibitem{pg}
Michael Powell, Se-Hyun Yang, Babak Falsafi, Kaushik Roy, and T.~N. Vijaykumar.
\newblock Gated-vdd: A circuit technique to reduce leakage in deep-submicron
  cache memories.
\newblock In {\em Proceedings of the 2000 International Symposium on Low Power
  Electronics and Design}, ISLPED '00, page 90–95, New York, NY, USA, 2000.
  Association for Computing Machinery.

\bibitem{srampg}
P.~{Nair}, S.~{Eratne}, and E.~{John}.
\newblock A quasi-power-gated low-leakage stable sram cell.
\newblock In {\em 2010 53rd IEEE International Midwest Symposium on Circuits
  and Systems}, pages 761--764, 2010.

\bibitem{noc1}
J.~Zhan, J.~Ouyang, F.~Ge, J.~Zhao, and Y.~Xie.
\newblock {DimNoC: A dim silicon approach towards power-efficient on-chip
  network}.
\newblock In {\em 2015 52nd ACM/EDAC/IEEE Design Automation Conference (DAC)},
  June 2015.

\bibitem{nvpg}
T.~{Ohsawa}, S.~{Ikeda}, T.~{Hanyu}, H.~{Ohno}, and T.~{Endoh}.
\newblock A 1-mb stt-mram with zero-array standby power and 1.5-ns quick
  wake-up by 8-b fine-grained power gating.
\newblock In {\em 2013 5th IEEE International Memory Workshop}, pages 80--83,
  2013.

\bibitem{mcrdram}
Jungwhan Choi, Wongyu Shin, Jaemin Jang, Jinwoong Suh, Yongkee Kwon, Youngsuk
  Moon, and Lee-Sup Kim.
\newblock Multiple clone row dram: A low latency and area optimized dram.
\newblock In {\em Proceedings of the 42nd Annual International Symposium on
  Computer Architecture}, ISCA '15, page 223–234, New York, NY, USA, 2015.
  Association for Computing Machinery.

\bibitem{clrdram}
Haocong Luo, Taha Shahroodi, Hasan Hassan, Minesh Patel, A.~Giray
  Ya\u{g}l\i{}k\c{c}\i{}, Lois Orosa, Jisung Park, and Onur Mutlu.
\newblock Clr-dram: A low-cost dram architecture enabling dynamic
  capacity-latency trade-off.
\newblock In {\em Proceedings of the ACM/IEEE 47th Annual International
  Symposium on Computer Architecture}, ISCA '20, page 666–679. IEEE Press,
  2020.

\bibitem{kang}
Sung-Mo~(Steve) Kang and Yusuf Leblebici.
\newblock {\em {CMOS Digital Integrated Circuits Analysis \& Design}}.
\newblock McGraw-Hill, Inc., 2003.

\bibitem{technode}
R.~{Teodorescu} and J.~{Torrellas}.
\newblock Variation-aware application scheduling and power management for chip
  multiprocessors.
\newblock In {\em 2008 International Symposium on Computer Architecture}, pages
  363--374, 2008.

\bibitem{liu-retention}
Jamie Liu, Ben Jaiyen, Yoongu Kim, Chris Wilkerson, and Onur Mutlu.
\newblock An experimental study of data retention behavior in modern dram
  devices: Implications for retention time profiling mechanisms.
\newblock {\em SIGARCH Comput. Archit. News}, 2013.

\bibitem{retention-paper-1}
Samira Khan, Donghyuk Lee, Yoongu Kim, Alaa~R. Alameldeen, Chris Wilkerson, and
  Onur Mutlu.
\newblock The efficacy of error mitigation techniques for dram retention
  failures: A comparative experimental study.
\newblock {\em SIGMETRICS Perform. Eval. Rev.}, 2014.

\bibitem{avatar}
M.~K. {Qureshi}, D.~{Kim}, S.~{Khan}, P.~J. {Nair}, and O.~{Mutlu}.
\newblock Avatar: A variable-retention-time (vrt) aware refresh for dram
  systems.
\newblock In {\em 2015 45th Annual IEEE/IFIP International Conference on
  Dependable Systems and Networks}, pages 427--437, 2015.

\bibitem{approx2}
Arnab Raha, Hrishikesh Jayakumar, Soubhagya Sutar, and Vijay Raghunathan.
\newblock Quality-aware data allocation in approximate dram*.
\newblock In {\em 2015 International Conference on Compilers, Architecture and
  Synthesis for Embedded Systems (CASES)}, pages 89--98, 2015.

\bibitem{rowclone}
Vivek Seshadri, Yoongu Kim, Chris Fallin, Donghyuk Lee, Rachata
  Ausavarungnirun, Gennady Pekhimenko, Yixin Luo, Onur Mutlu, Phillip~B.
  Gibbons, Michael~A. Kozuch, and Todd~C. Mowry.
\newblock Rowclone: Fast and energy-efficient in-dram bulk data copy and
  initialization.
\newblock In {\em Proceedings of the 46th Annual IEEE/ACM International
  Symposium on Microarchitecture}, MICRO-46, page 185–197, New York, NY, USA,
  2013. Association for Computing Machinery.

\bibitem{vsensors}
M.~{Meterelliyoz}, P.~{Song}, F.~{Stellari}, J.~P. {Kulkarni}, and K.~{Roy}.
\newblock A high sensitivity process variation sensor utilizing sub-threshold
  operation.
\newblock In {\em 2008 IEEE Custom Integrated Circuits Conference}, pages
  125--128, 2008.

\bibitem{trie_paper}
Baboescu Florin, Rajgopal Suresh, Huang Lun-Bin, and Richardson Nick.
\newblock {Hardware Implementation of a Tree Based IP Lookup Algorithm for
  OC-768 and beyond}.
\newblock Technical Report {ST Microelectronics Inc}, 2005.

\bibitem{dualedgeFF}
S.~H. {Unger}.
\newblock Double-edge-triggered flip-flops.
\newblock {\em IEEE Transactions on Computers}, C-30(6):447--451, 1981.

\bibitem{intelpaper}
{Intel}.
\newblock {Intel 64 and IA-32 Architectures Optimization Reference Manual}.
\newblock Technical report, April 2012.

\bibitem{gem5}
Nathan Binkert, Bradford Beckmann, Gabriel Black, Steven~K. Reinhardt, Ali
  Saidi, Arkaprava Basu, Joel Hestness, Derek~R. Hower, Tushar Krishna, Somayeh
  Sardashti, Rathijit Sen, Korey Sewell, Muhammad Shoaib, Nilay Vaish, Mark~D.
  Hill, and David~A. Wood.
\newblock {The Gem5 Simulator}.
\newblock {\em SIGARCH Comput. Archit. News}, 39(2), August 2011.

\bibitem{flikker}
Song Liu, Karthik Pattabiraman, Thomas Moscibroda, and Benjamin~G. Zorn.
\newblock Flikker: Saving dram refresh-power through critical data
  partitioning.
\newblock In {\em Proceedings of the Sixteenth International Conference on
  Architectural Support for Programming Languages and Operating Systems},
  ASPLOS XVI, page 213–224, New York, NY, USA, 2011. Association for
  Computing Machinery.

\bibitem{approx1}
Arnab Raha, Soubhagya Sutar, Hrishikesh Jayakumar, and Vijay Raghunathan.
\newblock Quality configurable approximate dram.
\newblock {\em IEEE Transactions on Computers}, 66(7):1172--1187, 2017.

\end{thebibliography}
